\documentclass[9pt,twoside]{article}

\usepackage{fullpage}
\usepackage{graphicx}
\usepackage{caption}
\usepackage{subcaption}
\usepackage{multirow}
\usepackage{verbatim}
\usepackage{amsmath}
\usepackage{amssymb}
\usepackage{color}
\usepackage{hyperref}
\usepackage{stackrel}
\usepackage{upgreek}
\usepackage{csquotes}

\def\green{\textcolor{green}}

\definecolor{lightgray}{gray}{0.8}
\definecolor{llightgray}{gray}{0.95}

\def\##1{{\underline #1}}
\def\=#1{\underline{\underline{#1}}}
\def\+#1{\underline{\bf #1}}
\def\*#1{\breve{\bf #1}}

\def\.{\mbox{ \tiny{$^\bullet$} }}

\def\le{\left(}
\def\ri{\right)}
\def\les{\left[}
\def\ris{\right]}
\def\lec{\left\{}
\def\ric{\right\}}

\def\r#1{(\ref{#1})}

\def\ko{k_{\scriptscriptstyle 0}}

\def\00{^{(0,0)}}

\def\ux{\hat{\#u}_{\rm x}}
\def\uy{\hat{\#u}_{\rm y}}
\def\uz{\hat{\#u}_{\rm z}}

\def\deg{^\circ}

\def\lambdao{\lambda_{\scriptscriptstyle 0}}

\def\lambdaomin{\lambda_{\scriptscriptstyle 0, {\rm min}}}
\def\lambdaomax{\lambda_{\scriptscriptstyle 0,{\rm max}}}    
\def\La{L_{\rm a}}	
\def\Ld{L_{\rm d}}	
\def\Lg{L_{\rm g}}	
\def\Lm{L_{\rm m}}

\def\LCIGS{L_{\rm CIGS}}	
\def\LCdS{L_{\rm CdS}}
\def\LZnO{L_{\rm iZnO}}
\def\Lt{L_{\rm t}}

\def\Lw{L_{\rm w}}
\def\Lx{L_{\rm x}}

\def\co{c_{\scriptscriptstyle 0}}

\def\eps{\varepsilon} 
\def\epso{\eps_{\scriptscriptstyle 0}}

\def\epsd{\eps_{\rm d}}
\def\epsm{\eps_{\rm m}}
\def\epsg{\eps_{\rm g}}
\def\epsdc{\varepsilon_{\rm dc}}

\def\sfE{{\sf E}}
\def\Eo{E_{\scriptscriptstyle 0}}
\def\Ei{\sfE_{\rm i}}
\def\Ec{\sfE_{\rm c}}
\def\Ev{\sfE_{\rm v}}
\def\eg{\sfE_{\rm g}}
\def\ego{\sfE_{\rm g,min}}

\def\egmax{\sfE_{\rm g,max}}
\def\EFn{\sfE_{\rm F_{n}}}
\def\EFp{\sfE_{\rm F_{p}}}

\def\etao{\eta_{\scriptscriptstyle 0}}

\def\Jsc{J_{\rm sc}}
\def\JscOpt{J_{\rm sc}^{\rm Opt}}

\def\Jn{J_{\rm n}}
\def\Jp{J_{\rm p}}

\def\Jdev{J_{\rm dev}}

\def\kB{k_{\rm B}}
\def\muo{\mu_{\scriptscriptstyle 0}}
\def\mun{\mu_{\rm n}}
\def\mup{\mu_{\rm p}}

\def\ni{n_{\rm i}}
\def\Nc{N_{\rm c}}
\def\Nv{N_{\rm v}}
\def\ND{N_{\rm D}}
\def\Nf{N_{\rm f}}

\def\vth{v_{\rm th}}

\def\Pmax{P_{\rm max}}
\def\Pin{P_{\rm in}}
\def\Psup{P_{\rm sup}}

\def\qe{q_{\rm e}}

\def\sigman{\sigma_{\rm n}}
\def\sigmap{\sigma_{\rm p}}

\def\Voc{V_{\rm oc}}

\def\Vext{V_{\rm ext}}

\def\Rnpz{R(n,p;z)}
\def\Rnpzrad{R_{\rm rad}(n,p;z)}
\def\RnpzSRH{R_{\rm SRH}(n,p;z)}
\def\RB{R_{\rm B}}

\def\taun{\tau_{\rm n}}
\def\taup{\tau_{\rm p}}
\def\Al2O3{\rm Al_2 O_3}
\def\LAl2O3{L_{\rm Al_2 O_3}}

\begin{document}

\begin{center}
{\Large{\textbf{Efficiency enhancement of  ultrathin  CIGS solar cells by optimal bandgap grading} }}\\

Faiz Ahmad$^1$, Tom H. Anderson$^2$, Peter B. Monk$^2$, and Akhlesh Lakhtakia$^1$\\

$^1${Pennsylvania State University, Department of Engineering Science and Mechanics, NanoMM--Nanoengineered Metamaterials Group,   University Park, PA 16802, USA}

$^2${University of Delaware, Department of Mathematical Sciences,
	501 Ewing Hall,  Newark, DE 19716, USA}

{Corresponding author: akhlesh@psu.edu}

\end{center}

\begin{abstract}
The power conversion efficiency of an ultrathin  CuIn$_{1-\xi}$Ga$_{\xi}$Se$_2$   (CIGS) solar cell was maximized  using a coupled optoelectronic model to determine the optimal bandgap grading
of the  nonhomogeneous CIGS layer  in the thickness direction. The bandgap of the CIGS layer was either sinusoidally or linearly graded, and the solar cell was modeled to
have  a  metallic  backreflector corrugated periodically along a fixed direction in the plane.  The model
predicts that
specially tailored bandgap grading can significantly improve the efficiency, with much smaller improvements due to the periodic corrugations. An efficiency of $27.7$\% with the conventional 2200-nm-thick CIGS layer is predicted with sinusoidal bandgap grading, in comparison to 22\% efficiency obtained experimentally with homogeneous bandgap.
Furthermore, the inclusion of sinusoidal grading increases the predicted efficiency to 22.89\% with just a 600-nm-thick CIGS layer.  These high efficiencies arise due to a large electron-hole-pair generation rate in the narrow-bandgap regions and the elevation of the open-circuit voltage due to a wider bandgap in the region toward the front surface of the CIGS layer.  
Thus, bandgap nonhomogeneity, in conjunction with periodic corrugation of the backreflector, can be effective in realizing ultrathin  CIGS solar cells  that can  help overcome the scarcity of indium.
	
\end{abstract}

 	\def\doubleunderline#1{\underline{\underline{#1}}}
	\renewcommand\vec{\mathbf}

\section{Introduction}
The power conversion efficiency $\eta$ of
thin-film CuIn$_{1-\xi}$Ga$_{\xi}$Se$_2$ (CIGS) solar cells is about $22\%$~\cite{ZWS,Green2018}. 
This efficiency  compares well to the 26.7\% efficiency of single-junction crystalline-silicon solar cells \cite{Green2018}, but the scarcity of indium  is a major obstacle 
for large-scale and low-cost production of   thin-film CIGS
solar cells\cite{Candelise_et_al-2012}. If the thickness of the CIGS   layer could  be reduced without significantly reducing the efficiency, this obstacle could be overcome. However, na\"ively reducing that  thickness below
1000~nm can lower $\eta$ for two reasons. First, the lower 
absorption of solar photons  would reduce both the optical short-circuit current density $\JscOpt$ and the  open-circuit voltage $\Voc$. Second, the  increased back-contact electron-hole-pair 
recombination rate~\cite{Gloeckler-Sites2005, Schmid2017} would reduce the output current density. Common techniques under investigation to offset these problems
include the use of light-trapping nanostructures  \cite{Schmid2017, Schmid2015, P-Lalanne2017},  alternative back contacts~\cite{P-Lalanne2017}, and back-surface passivation~\cite{Vermang}. 
	
Although nonhomogeneity (i.e., bandgap grading) of the CIGS  layer could increase $\eta$ by establishing drift fields~\cite{Gloeckler-Sites2005}, simple simulations as well as experiments have shown that linear grading of the bandgap can significantly reduce the short-circuit current density $\Jsc$~\cite{Gloeckler-Sites2005, Gloeckler-Sites2005JPCS, Songetal2004, Song2010}.
New strategies are required for bandgap grading to maintain $\Jsc$ and  enhance $\Voc$.   
	
Optoelectronic simulations of thin-film Schottky-barrier solar cells with periodically nonhomogeneous absorbing layers of InGaAn and   periodically corrugated backreflectors, predict improvement in efficiency ~\cite{Anderson2017,Anderson2018}. 
	This improvement is due to: (i) the enhancement of $\Jsc$ owing to the excitation of guided wave modes by the use of the periodically corrugated backreflector~\cite{Faryad2013, Liu2015, Haug2011, Khaleque2013} and
	(ii) the enhancement of $\Voc$ because of bandgap grading in the nonhomogeneous semiconductor layer~\cite{Gloeckler-Sites2005, Song2010, Songetal2004}.  Motivated by these simulations,
	we undertook a detailed optoelectronic optimization  of   ultrathin CIGS solar cells   with a 
	nonhomogeneous CIGS   layer with back-surface passivation and 
	backed by a periodically corrugated metallic backreflector.
	
Nonhomogeneity in the CIGS   layer  was modeled through either a sinusoidal or a linear variation of the bandgap along the thickness direction (taken to be the $z$ axis
of a Cartesian coordinate system). The commonly considered planar molybdenum (Mo) contact, which also functions as a backreflector, 
was  {taken to be periodically corrugated
along the $x$ axis for better light trapping~\cite{Faryad2013, Haug2011, Khaleque2013}.  
Parenthetically, the replacement of Mo by silver (Ag) has been theoretically predicted to enhance
$\JscOpt$~\cite{P-Lalanne2017}, but the lower stability of Ag at temperatures exceeding
550~$\deg$C makes it impractical for the fabrication of CIGS solar cells.   
A thin passivation layer of $\Al2O3$} was inserted between the CIGS layer and the backreflector. This passivation layer 
reduces the back-contact electron-hole recombination rate~\cite{Vermang} and also
protects the electrical characteristics of the CIGS   layer~\cite{Fonash}. 
	
We first used the rigorous
coupled-wave approach (RCWA)~\cite{GG, ESW2013} to calculate the
useful absorptance \cite{Ahmad2018}  of the chosen solar cell exposed to
normally incident unpolarized light. Assuming the incident power spectrum was the AM1.5G solar spectrum \cite{SSAM15G}, 
we then determined the $x$-averaged electron-hole-pair generation rate $G(z)$ in the solar cell. Then, we implemented
the one-dimensional (1D) drift-diffusion model~\cite{Fonash,Jenny_Book}  
for electrical calculations. A hybridizable discontinuous Galerkin (HDG) scheme~\cite{Lehrenfeld, CockburnHDG} was developed for the drift-diffusion equations. 
The bandgap-dependent electron affinity and defect density~\cite{Frisk14} were incorporated in the electrical calculations, as also were
the nonlinear Shockley--Read--Hall (SRH) and radiative electron-hole recombination processes~\cite{Fonash, Jenny_Book}.  
{Surface recombination was neglected because it has been experimentally shown
to be inconsequential for CIGS solar cells \cite{Kuciauskas2013}, but we did assess
the role of   traps at a  
CdS/CIGS interface in the solar cell \cite{Frisk14} by incorporating 
a surface-defect layer
\cite{Songetal2004}.}   
The optoelectronic model was implemented for 
the conventional  2200-nm-thick homogeneous CIGS layer for  ${\xi}\in\left\{0, 0.25, 1\right\}$ and
the predicted efficiency compared favorably with   experimental results in the literature.

The bandgap profile of the CIGS layer and 
the dimensions of the backreflector
were optimized for discrete values of
the thickness $\LCIGS$  of the CIGS   layer ranging from  {100~nm to 2200~nm.} The differential evolution algorithm
(DEA)~\cite{DEA} was used to maximize $\eta$ for three configurations:
\begin{itemize}
\item[(i)] homogeneous-bandgap CIGS layer with {flat   backreflector},
\item[(ii)] sinusoidally nonhomogeneous-bandgap CIGS layer
with periodically corrugated backreflector, and 
\item[(iii)] linearly nonhomogeneous-bandgap with periodically corrugated backreflector.  
\end{itemize}

By implementing a coupled optoelectronic model instead of an optical model, we were
able to bypass the major limitation of the latter: optical models can yield $\JscOpt$ 
as well as  the theoretical upper bound $\Psup$ of the maximum extractable power density ~\cite{Ben2018}, 
but are incapable of accurately modeling  $V_{oc}$ and   $\eta$. Not surprisingly therefore, optical models 
yield homogeneous bandgap to be superior to bandgap grading for maximizing  $\Psup$ \cite{Ahmad-SPIE2018}, 
but that conclusion is irrelevant for the maximization of $\eta$, as we show in this paper.	
	
The structure of this paper is as follows.  Section~\ref{Opt-optimiz} on optoelectronic optimization is 
divided into five subsections.
The optical description of the solar cell is presented in Sec~2.\ref{sec:geom}
and the approach adopted for optical calculations is
summarized in Sec.~2.\ref{sec:optical-theory}, the electrical description of
the solar cell is discussed in Sec~2.\ref{sec:EDSC} and the equations solved
are described in Sec.~2.\ref{sec:ElecModel}, and optimization is discussed in Sec.~2.\ref{sec:JVEF}.
 
Numerical results are presented and discussed in Sec.~\ref{sec:OptoElecRes},
which is  divided into seven subsections. 
 {Section~3.\ref{sec:Ref-conventional-cell} compares the efficiency of
the   conventional 2200-nm-thick solar cell with a homogeneous  CIGS
layer predicted by the model with available experimental data.  
The effect of the $\Al2O3$ passivation layer on the solar-cell performance is discussed
in Sec.~3.\ref{Al203_passivation}.
Section~3.\ref{sec:opto_elechomo_FBR} provides
the optimal results for solar cells with a homogeneous  CIGS
layer and a flat backreflector,
Sec.~3.\ref{sec:opto_elechomo_PCBR}  for  solar cells with a homogeneous  CIGS
layer and a periodically corrugated backreflector, and
Sec.~3.\ref{sec:Linearly_nonhomo}  for solar cells
with a linearly graded CIGS layer and a periodically corrugated backreflector. 
Optimal results for solar cells with a   sinusoidally graded CIGS layer and a periodically
corrugated backreflector are discussed in Sec.~3.\ref{sec:optoelec_nonhomo_PCBR}. 
A detailed study of the optimal 600-nm-thick solar cell is presented in Sec.~3.\ref{sec:Optimal_design}.
The paper ends with concluding remarks in Sec.~\ref{sec:conc}. }
	
All optical calculations were performed with an implicit $\exp(-i\omega{t})$ dependence on time $t$,
with $\omega$ as the angular frequency and $i=\sqrt{-1}$.
The free-space wavelength and the intrinsic impedance of free space are denoted by $\lambdao=2\pi\co/ \omega$ and $\etao=\sqrt{\muo/\epso}$, respectively,  where  $\lambdao$ is the free-space wavelength, 
$\muo$   is  the permeability of free space, 
$\epso$ is  the   permittivity of free space,
and $\co=1/\sqrt{\epso\muo}$ is the speed of light in free space.  
Vectors are underlined and the Cartesian unit vectors are identified as $\ux$, $\uy$, and $\uz$.

\section{Optoelectronic optimization}\label{Opt-optimiz}
\begin{figure}[htb]
		\centering	
		\includegraphics[width=0.5\textwidth]{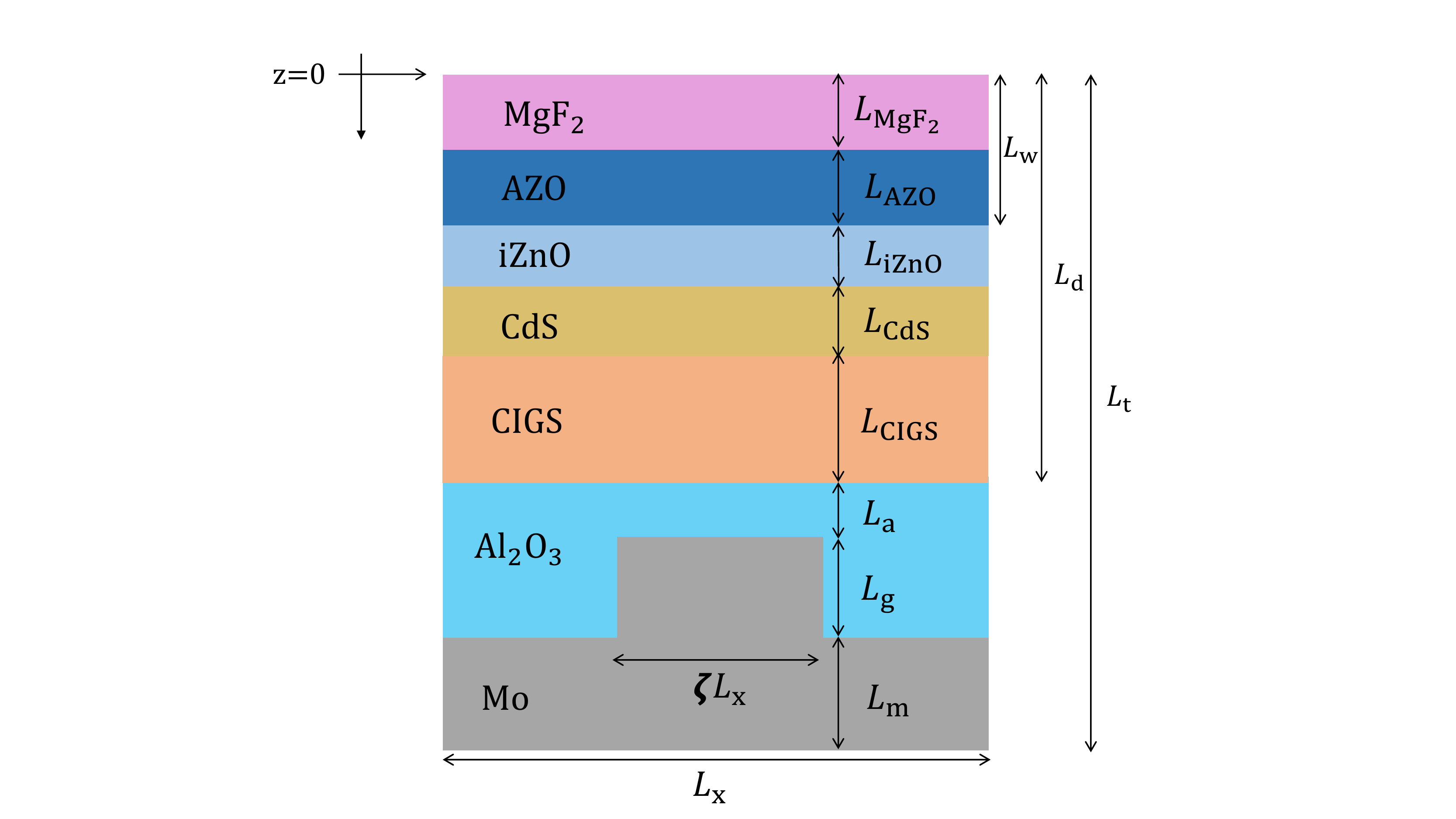} 
		\caption{Schematic of the reference unit cell of the CIGS solar cell with a 1D periodically corrugated metallic backreflector.} \label{figure1} 
	\end{figure}

The structure of the CIGS solar cell is shown schematically in  Fig.~\ref{figure1}. The solar cell 
occupies the region 
${\cal X}:\left\{(x,y,z)
\vert -\infty<x<\infty, -\infty<y<\infty, 0<z<\Lt\right\}$,
with the half spaces $z<0$ and $z>\Lt$ occupied by air. The reference unit cell of this structure is 
identified as 	${\cal R}:\left\{(x,y,z)
\vert -\Lx/2<x<\Lx/2,  -\infty<y<\infty,  0<z<\Lt\right\}$,
with the backreflector being  periodically corrugated with period $\Lx$ along the $x$ axis.

The window region $0<z<\Lw=210$~nm consists of a
 110-nm-thick layer of magnesium fluoride (MgF$_2$)~\cite{mgf2}  as an antireflection coating~\cite{Rajan} and a 100-nm-thick layer
 of aluminum-doped zinc oxide (AZO) \cite{AZO}  as an electrical contact.
 The region $\Lw<z<\Lw+\LZnO$ is a 80-nm-thick layer of intrinsic 
 zinc oxide (iZnO) \cite{iZnO} as a buffer layer to increase the open-circuit voltage
 \cite{Jahagirdar2003}. The region $\Lw+\LZnO<z<\Lw+\LZnO+\LCdS$ is a 70-nm-thick layer of $n$-type CdS \cite{treharne} to form a junction with a $p$-type CIGS layer of thickness $\LCIGS= \Ld-\Lw-\LZnO-\LCdS
 \in\lec 100, 200, 300, 400, 500, 600,900,1200,2200\ric$~nm. 
	
The region $\Ld<z<\Ld+\La$ of thickness $\La=50$~nm is occupied by $\Al2O3$ \cite{Al2O3}. This is included to protect the electrical characteristics of the CIGS layer and also function as a passivation layer to reduce the back-surface electron-hole recombination rate. The  region $\Ld+\La+\Lg<z<\Lt$
is occupied by  Mo \cite{Mo}, the thickness $\Lm=\Lt-\left(\Ld+\La+\Lg\right)=500$~nm
being  chosen to be well beyond the electromagnetic penetration depth \cite{Iskander} of Mo in the optical regime.
The region $\Ld+\La<z<\Ld+\La+\Lg$ consists of a  rectangular 
Mo grating with period $\Lx$ along the $x$ axis.
	
\subsection{Optical description of solar cell}\label{sec:geom}
   
\begin{figure}[htb]
\centering
\includegraphics[width=0.5 \textwidth]{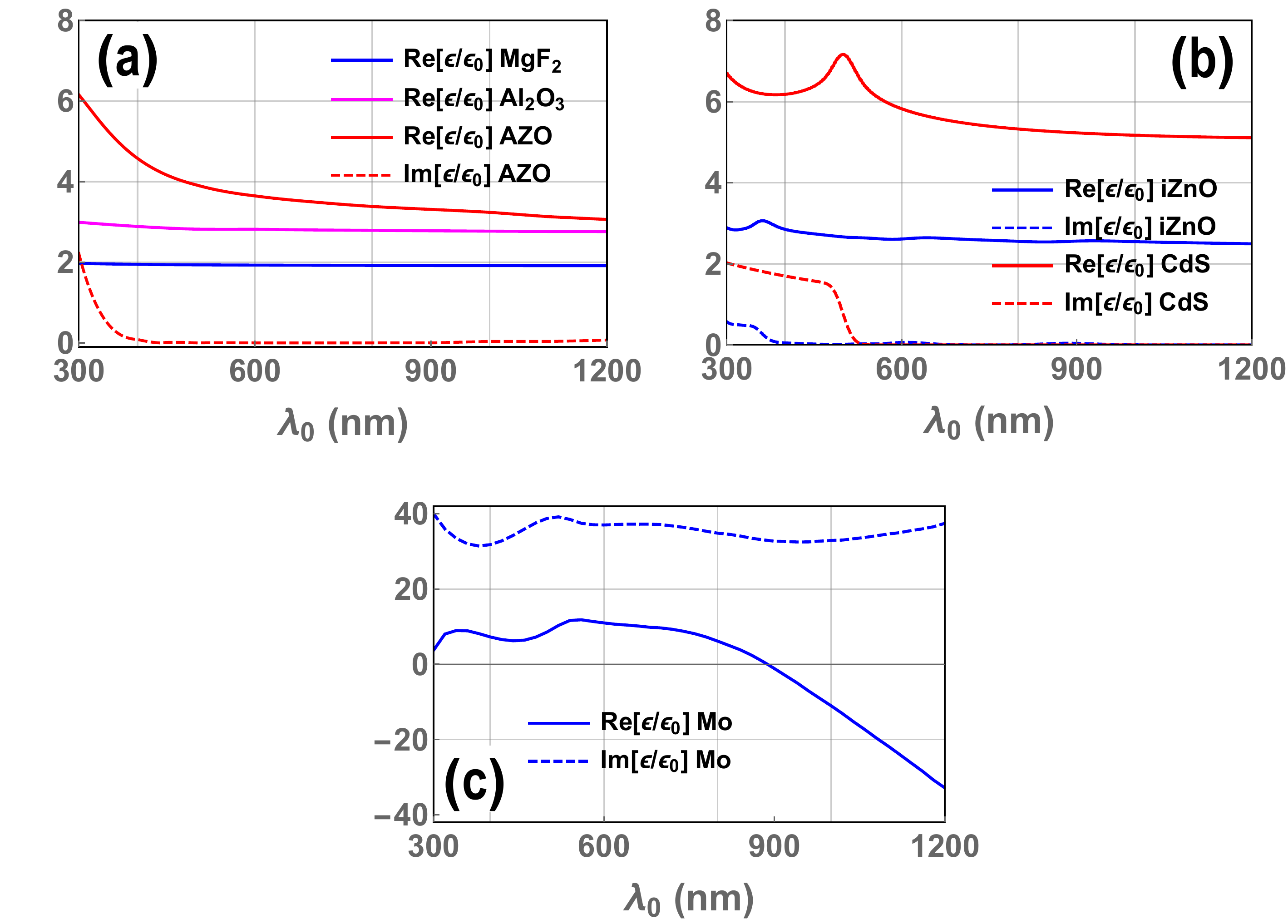}    
\caption{(a) Real and  imaginary parts of the relative permittivity $\eps/\epso$
 of MgF$_2$, $\Al2O3$, and AZO  as  functions of $\lambdao$. The
 imaginary part of the relative permittivity of MgF$_2$ is negligibly
 small. The imaginary part of the relative permittivity of  $\Al2O3$ is zero. (b) Real and  imaginary parts of the relative permittivity $\eps/\epso$ of 
 iZnO  and CdS as functions of $\lambdao$.
 (c) Real and  imaginary parts of the relative permittivity $\eps/\epso$ of 
 Mo as functions of $\lambdao$.
 \label{spectra-eps-1}}
\end{figure}

The permittivity in the grating region of $\cal R$ is given by
\begin{eqnarray}
\nonumber
 &&
\epsg(x, z,\lambdao) =\left\{\begin{array}{ll}\epsm(\lambdao) \,,&\qquad \vert{x}\vert<\zeta \Lx/2\,,
\\[4pt] \epsd(\lambdao) \,,& 
\qquad \vert{x}\vert>\zeta \Lx/2\,,
\end{array}\right.
\\
&&		
\quad z\in(\Ld+\La,\Ld+\La+\Lg)\,,
\label{Eqn:Grating}
\end{eqnarray}
where $\zeta\in[0,1]$ is the duty cycle, $\epsm(\lambdao)$ is the permittivity of Mo {\cite{Mo},}
and $\epsd(\lambdao)$ is the permittivity of $\Al2O3$ \cite{AZO}. 		
{Spectrums} of real and imaginary parts of the relative permittivity
$\eps(\lambdao)/\epso$ of MgF$_2$~\cite{mgf2}, AZO~\cite{AZO}, iZnO~\cite{iZnO}, CdS~\cite{treharne}, $\Al2O3$ \cite{Al2O3},  and Mo \cite{Mo}   used in our 
calculations are displayed in Fig.~\ref{spectra-eps-1}.  
		
The bandgap $\eg$ of CIGS  varies  with $z$ in the CIGS layer. As solar cells are fabricated using vapor-deposition techniques \cite{RJMP}, nonhomogeneous bandgap profiles could be physically realized by varying  the parameter  $\xi \in[0,1]$   during the deposition process \cite{Frisk14, Lindahl2013}. 
The linearly nonhomogeneous bandgap for forward  grading was modeled as 
	\begin{eqnarray} 
	\nonumber
	&&\eg(z)=\ego 
	\\ &&+A\left(\egmax-\ego\right)\frac{z-\left(\Lw+\LZnO+\LCdS\right)}{\LCIGS}\, ,
	\nonumber
	\\
	&&
	\qquad z\in\left[\Lw+ \LZnO+\LCdS, \Ld\right]\, ,
	\label{Eqn:Linear-bandgap}
	\end{eqnarray}
where   $\ego$ is the minimum
bandgap, $\egmax$ is the maximum bandgap,
and $A$ is an amplitude (with $A = 0$ representing a homogeneous CIGS layer). The   linearly nonhomogeneous
bandgap for backward
 grading was modeled as
	\begin{eqnarray}
	\nonumber
	&&
	\eg(z)=\egmax 
	\\&&-A\left(\egmax-\ego\right)\frac{z-\left(\Lw+\LZnO+\LCdS\right)}{\LCIGS}\, , 
	\nonumber
	\\ &&\qquad z\in\left[\Lw+\LZnO+\LCdS, \Ld\right]\, .
	\label{Eqn:Linear-bandgap1}
	\end{eqnarray} 
Three representative profiles of linearly  nonhomogeneous bandgap are shown in  Fig.~\ref{CIGS-profile}(a). The parameter space for optimization of $\eta$ was fixed as
follows: {$\Lg\in\les0, 550\ris~\rm nm$, 
		$\zeta\in\les0, 1\ris$, $\Lx\in\les100, 1000\ris~\rm nm$, $A\in\les0, 1\ris$,
		$\ego\in\les 0.947, 1.626\ris~\rm$~eV, and  $\egmax\in\les 0.947, 1.626\ris$~eV with the condition $\egmax \geq \ego$.

The sinusoidally varying bandgap was modeled as
\begin{align} 
\nonumber
&\eg(z)=\ego +A\le 1.626-\ego\ri \, 
\\
\nonumber
&\times
\lec \frac{1}{2}\, \les \sin\le 2\pi K \frac{z-\left(\Lw+\LZnO+\LCdS\right)}{\LCIGS}-2\pi \psi\ri \, +1\ris\, \ric^{\alpha} \, , 
\\[5pt]
&\qquad\qquad z\in\left[\Lw+\LZnO+\LCdS, \Ld\right]\, ,
\label{Eqn:Sin-bandgap}
\end{align}
where  $\psi\in [0, 1)$
quantifies a relative phase shift, $K$ is the number of periods in the CIGS layer, and $\alpha> 0 $ is a
shaping parameter. Three representative profiles of sinusoidally  nonhomogeneous bandgap are shown in Fig.~\ref{CIGS-profile}(b). The parameter space for optimization of $\eta$  was fixed as
follows: $\Lg\in\les0, 550\ris~\rm nm$, $\zeta\in\les0, 1\ris$, $\Lx\in\les100, 1000\ris~\rm nm$,  $A\in\les0, 1\ris$, $\ego\in\les 0.947, 1.626\ris~\rm eV$, 
$\alpha\in\les0, 7\ris$, $K\in\les0, 8\ris$, and $\psi\in\les0,1\ris$.

Spectrums of the real and imaginary parts of the relative permittivity $\eps/\epso$ of CIGS in the optical regime  are plotted in Fig.~\ref{figure4}
as   functions of $\xi$~\cite{Ahmad-SPIE2018,Minoura}.

\begin{figure}[htb]
\centering	
\includegraphics[width=0.5\textwidth]{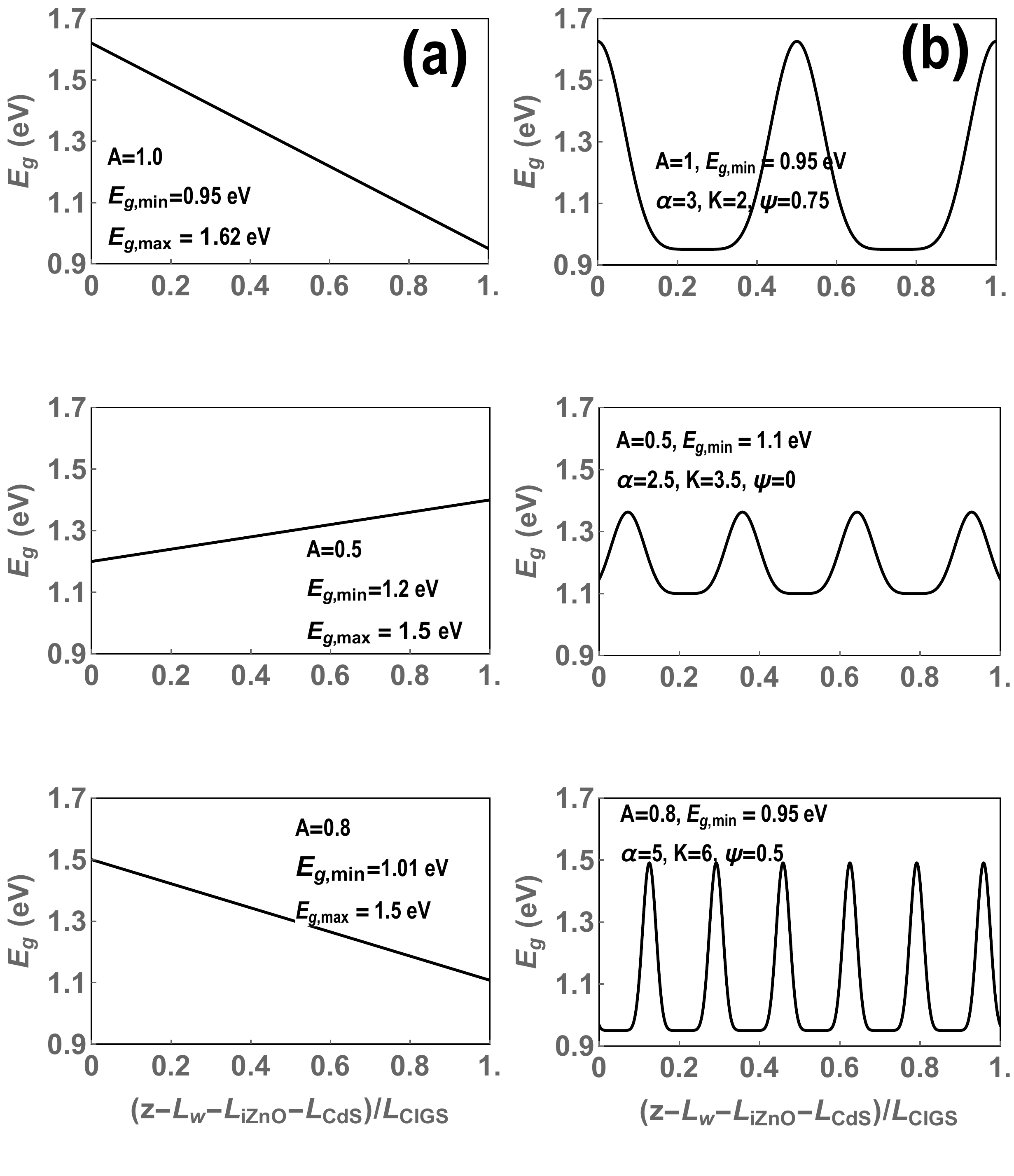} 
\caption{\label{CIGS-profile}(a) Three representative profiles of the linearly nonhomogeneous  bandgap of the CIGS layer.
(b) Three representative profiles of the sinusoidally nonhomogeneous  bandgap of the CIGS layer.} 
\end{figure}

\begin{figure}[htb]
\centering
\includegraphics[width=0.6\textwidth]{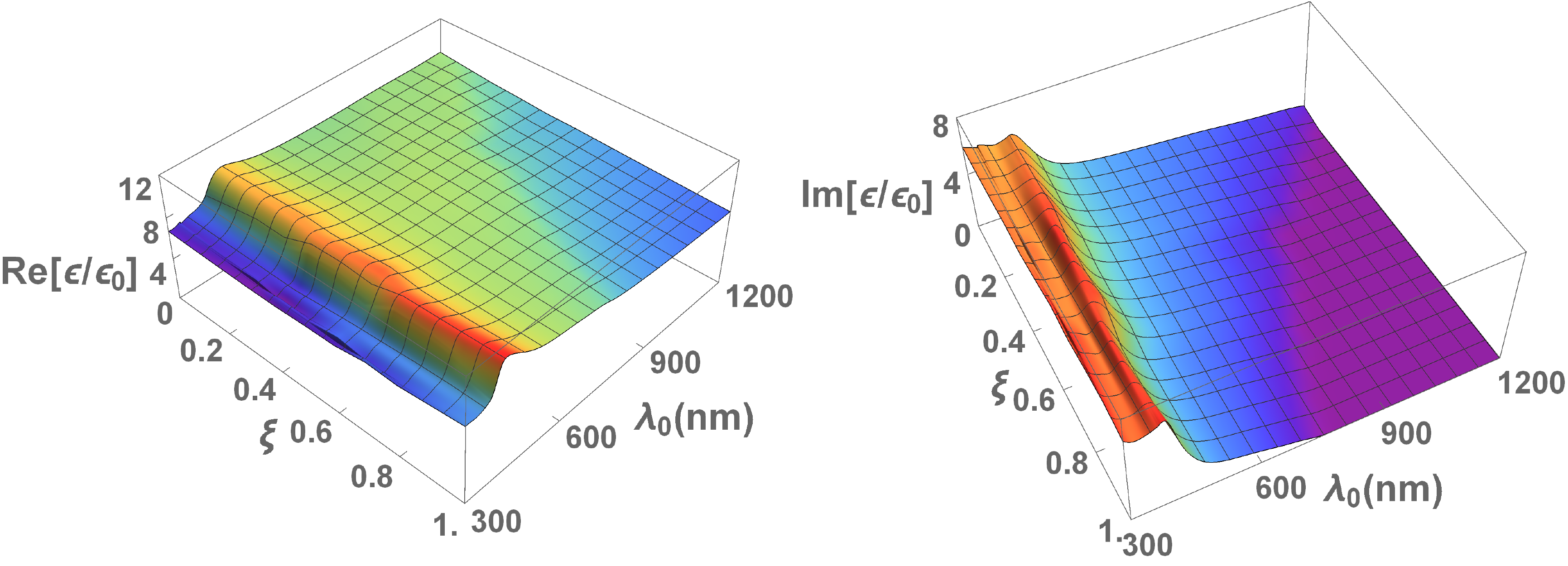}    
\caption{Real and  imaginary parts of the relative permittivity $\eps/\epso$ of CIGS as functions
of $\lambdao$ and  $\xi$.
\label{figure4}}
\end{figure}
		
\subsection{Optical calculations} \label{sec:optical-theory}
The RCWA ~\cite{GG, ESW2013} was used the calculate the electric field phasor ${\#E}(x,z,\lambdao)$ everywhere inside the solar cell as a result of illumination by a monochromatic plane wave normally incident on the plane $z=0$ from the half space $z<0$. The electric field phasor of the incident plane wave was taken as
 \begin{equation}
 {\#E}_{\rm inc}(z,\lambdao)=
 {\Eo} \frac{\ux+\uy}{\sqrt{2}}\exp\left(i\ko{z}\right)\,,
 \end{equation}
where $\Eo=4\sqrt{15 \pi}$~V~m$^{-1}$. The region $\cal R$ was partitioned into a sufficiently large number of slices along the $z$ axis, such that the useful solar absorptance \cite{Ahmad2018, Ahmad-SPIE2018} converged correct to $\pm1\%$. Each  slice was taken to be homogeneous along the $z$ axis but {could} be periodically nonhomogeneous along the $x$ axis. Standard boundary conditions were enforced on the planes $z=0$ and $z=\Lt$ to match the internal field phasors to the incident, reflected, and transmitted  field phasors, as appropriate. Detailed descriptions of the RCWA implementation  are available elsewhere~\cite{Ahmad-SPIE2018, ESW2013, Anderson2018}.  
		
With the assumption that every absorbed photon excites an electron-hole pair,
the electron-hole-pair generation rate   was calculated as 			
\begin{equation}	
G_{2D}(x,z)=\frac{\etao}{\hbar\Eo^2 }\int_{\lambdaomin}^{\lambdaomax} {\rm Im}\{\eps(x, z, \lambdao)\}
\left\vert\#E(x, z, \lambdao)\right\vert^2\,S(\lambdao)  \,  d\lambdao
\label{G2D-def}
\end{equation}	
for
$z\in\les \Lw, \Ld \ris$,
where  $\hbar$ is  the reduced Planck constant,
$S(\lambdao)$ is the AM1.5G solar spectrum~\cite{SSAM15G},
$\lambdaomin = 300$~nm, and  $\lambdaomax = \left(1240/\ego\right)$~eV~nm.
As the solar cell operates under the influence of a $z$-directed electrostatic field created by the
application of a bias voltage $\Vext$,  charge carriers  generally flow along the $z$ axis, any current generated parallel to the $x$ axis being very small. Moreover, the period $\Lx$ of the corrugated backreflector  is  $\sim$500~nm, 
which is so small in comparison to the lateral dimensions of the solar cell that it can be
ignored for electrostatic analysis. Therefore,
the $x$-averaged  electron-hole-pair generation rate   was calculated as 			
\begin{equation}		
G(z)=\frac{1}{\Lx}  \int_{-\Lx/2}^{\Lx/2}\,  G_{2D}(x,z)\,dx  \, , \qquad z\in\les \Lw, \Ld \ris \,
\label{G1D-def}
\end{equation}	
for use in Secs.~2.\ref{sec:EDSC} and 2.\ref{sec:ElecModel}.  The generation rate $G (z)$ contains the effects of: (i) the periodic corrugations of the backreflector, (ii) the $\Al2O3$ back-surface passivation layer,  and (iii) the MgF$_2$   antireflection coating.

\subsection{Electrical description of solar cell}\label{sec:EDSC}
The region $\Lw<z<\Ld$ containing the iZnO, CdS, and CIGS layers was considered for electrical modeling.  Both iZnO and CdS must be considered in addition to CIGS, because both
contribute to charge-carrier generation. The useful solar spectrum is typically taken to span
free-space wavelengths in excess of $\lambdaomin = 300$~nm. Since iZnO has a bandgap of
3.3~eV, it will absorb solar photons with energies corresponding
to $\lambdao\in\les 300,376\ris$~nm. Likewise, as CdS has a
bandgap of 2.4~eV, it will absorb solar photons with energies corresponding
to $\lambdao\in\les 300,517\ris$~nm. Hence,
the generation of electron-hole pairs in the iZnO and CdS layers must be accounted for,
not to mention the recombination of electron-hole pairs in both layers.

As our focus  is on modeling the electrical characteristics of the
solar cell, not on how it interfaces with an external circuit, both terminals were assumed to
be ideal ohmic contacts. We used a 1D drift-diffusion model~\cite{Jenny_Book, Fonash, Brezzi2002} to investigate
the transport of electrons and holes for $z\in\left[\Lw,\Ld\right]$, as discussed next.
		
\subsection{Electrical theory}\label{sec:ElecModel}
The electron-current density $\Jn(z)$ and 
the hole-current density $\Jp(z)$ are driven by gradients in the electron and hole quasi-Fermi levels, respectively. Thus \cite[Sec.~4.6]{Jenny_Book},
\begin{equation}
\left.\begin{array}{l}
\Jn(z) = \displaystyle{\mun\, n(z)  \frac{d}{dz}\EFn(z) }
\\[5pt]
\Jp(z) = \displaystyle{\mup\, p(z)  \frac{d}{dz}\EFp(z)}
\end{array}\right\}\,,
\quad
z\in(\Lw,\Ld)
\label{Eqn:JnJp},
\end{equation}
where $\qe = 1.6\times$10$^{-19}$~C is the elementary charge;
$n(z)$ and $p(z)$ are   the   electron density and hole density, respectively; and
$\mun$ and $\mup$  are  the electron mobility and hole mobility, respectively. The electron   
quasi-Fermi level
\begin{equation}
\EFn(z)=\Ec(z)+\left(\kB T\right)\ln \les n(z)/\Nc(z)\ris
\label{EFn-def}
\end{equation}
and the hole quasi-Fermi level
\begin{equation}
\EFp(z)=\Ev(z)-\left(\kB T\right)\ln \les p(z)/\Nv(z)\ris
\label{EFp-def}
\end{equation}
involve the product of the Boltzmann constant $\kB = 1.380649 \times 10^{-23}$~J~K$^{-1}$
and the absolute temperature $T$. Furthermore,
$\Nc(z)$ is the density of states in the
 conduction band, $\Nv(z)$ is the density of states in the valence band,
 $\Ec(z)={\sf E}_0-\les \qe\phi(z)+\chi(z)\ris$  is the conduction band-edge energy,
 $ \Ev(z)=\Ec(z)-\eg(z)$ is the valence band-edge energy,  $\phi(z)$ is 
 the dc electric potential, and $\chi(z)$ is the
 bandgap-dependent electron affinity.   The reference energy level ${\sf E}_0$ is arbitary.
 When the  right side of Eq.~(\ref{EFn-def}) is substituted
 in Eq.~(\ref{Eqn:JnJp})$_1$, the contribution of diffusion of electrons to $\Jn(z)$ can be identified as depending on  $dn/dz$, the
 remainder being the contribution of electron drift; and likewise for $\Jp(z)$.

According to the Boltzmann approximation~\cite{Jenny_Book},  
\begin{equation}
\left.\begin{array}{l}
n(z) = \displaystyle{ \ni(z) \exp\lec\les{\EFn (z)- \Ei(z)}\ris/{\kB T}\ric}
\\
p(z) = \displaystyle{\ni(z)\exp\lec-\les{\EFp(z) - \Ei(z)}\ris/{\kB T}\ric}
\end{array}\right\}\,,
\label{Eqn:ndef-pdef}
\end{equation}
where  the intrinsic charge-carrier density  
		\begin{equation}
		\ni(z) = \sqrt{\Nc(z) \,\Nv(z)\, \exp\les- {\eg(z)}/{\kB T}\ris}
		\end{equation}
and the intrinsic energy
		\begin{equation}
		\Ei(z)=(1/2)\lec\Ec(z)+\Ev(z) {-\left(\kB T\right)}\ln\les\Nc(z)/\Nv(z)\ris\ric\,.
		\end{equation}

Under steady-state conditions,
the 1D drift-diffusion model  comprises the following three differential equations \cite[Sec.~4.6]{Jenny_Book}:
\begin{align}
\frac{d}{dz}\Jn(z)&= -\qe\les G(z) - \Rnpz\ris\label{Eqn:dJn},\\
\frac{d}{dz}\Jp(z)&= \qe\les G(z) - \Rnpz\ris\label{Eqn:dJp},\\
\epso\frac{d}{dz}\les\epsdc(z)  \frac{d}{dz}\phi(z)\ris&=
- \qe\les\Nf(z) +  \ND(z)+ p(z) - n(z)\ris. 
\label{Eqn:Poisson}
\end{align}
These differential equations hold for $z\in(\Lw,\Ld)$, with
$\Rnpz$  as the electron-hole-pair recombination rate,
 $\Nf(z)$ as  the defect density {(also called trap density)}, 
$\ND(z)$ as the donor density which is positive for donors and negative for acceptors,
 and 
$\epsdc (z)$   as the dc relative permittivity.
Although $\epsdc (z)$,
 $\Nc(z)$, and $\Nv(z)$  depend on $z$ because they depend
on the bandgap, following Frisk \text{et al.}~\cite{Frisk14} we took all three
quantities to be independent of $z$ because bandgap-dependent values
are not available for CIGS. 
		
Electron-hole pairs are produced at the rate $G (z)$ 
defined in Eq.~\r{G1D-def} and  recombine at the rate $\Rnpz$. 
We incorporated the radiative recombination 
and SRH recombination processes via $\Rnpz =\Rnpzrad+\RnpzSRH$ \cite{Jenny_Book, Fonash}. 
The radiative recombination rate is given by
		\begin{equation}
		\Rnpzrad = \RB \les n(z)p(z) - \ni^2(z)\ris,\label{rad}
		\end{equation}
where $\RB$ is the radiative recombination coefficient. 
The  SRH recombination rate is given by
		\begin{equation}
		\RnpzSRH  = \frac{n(z)p(z) - \ni^2(z)}{\taup(z)\les n(z) + n_1(z)\ris + \taun(z)\les p(z)+p_1(z)\ris}\,,\label{SRH}
		\end{equation}
where $n_1(z)$ and $p_1(z)$ are the electron and hole densities  at the trap energy level $\sfE_{\rm T}$; the minority carrier lifetimes
		\begin{equation}
		\tau_{\rm n,p}(z)=\displaystyle{{1}/{\les\sigma_{\rm n,p} \vth \Nf(z)\ris}}
		\end{equation}
depend on   the capture cross sections $\sigman$ and $\sigmap$ for electrons and holes, respectively; and
$v_{\rm th}$ represents the mean thermal {speed} for all charge carriers. The defect density $\Nf(z)$ was
taken to be bandgap-dependent for the SRH recombination rate.  
Equation~(\ref{Eqn:Poisson}) describes the dc electric potential created by the electrically charged regions of the solar cell.
Table~\ref{tab--elec-prop} provides the values of {the aforementioned}
electrical parameters {as well as the doping density $\ND$} used for iZnO, CdS, and CIGS~\cite{Frisk14}.

Equations~\r{Eqn:dJn}--\r{Eqn:Poisson} were supplemented by Dirichlet boundary conditions on $n(z)$, $p(z)$, and
$\phi(z)$ at the planes $z=\Lw$ and $z=\Ld$ \cite{Anderson2018,Anderson-JCP}. These boundary conditions were derived after assuming
the region $\Lw<z<\Ld$ to be uncharged and at local quasi-thermal equilibrium~\cite{Fonash};
furthermore, a  bias voltage $\Vext$ was applied at the plane $z=\Ld$. Solution of
this system of equations was undertaken using the HDG scheme~\cite{Chen2016, CockburnHDG, FuQiuHDG}, in which all the
$z$-dependent variables have to be discretized using discontinuous finite elements in a space of piecewise polynomials
\green{of} a fixed degree.
The full discretized system was solved for $n(z)$, $p(z)$, and $\phi(z)$,
using the Newton--Raphson method \cite{Brezzi2002}. The HDG scheme is particularly
advantageous for simulating solar cells with heterojunction interfaces~\cite{Brinkman} such as those which occur between
the CdS and CIGS layers.
		
\begin{table}[htb] 
	\caption{Electrical parameters of iZnO, CdS, and CIGS~\cite{Frisk14}.\label{tab--elec-prop}}
	\begin{tabular}{|p{1.8cm}|p{1.4cm}|p{1.4cm}|p{2.6cm}|} \hline \hline
		Parameter (unit)&iZnO & CdS&CIGS\\\hline
		$\eg$~(eV) &3.3&$2.4$ & $0.947$--$1.626$\, (Ga-dependent)\\ \hline
		$\chi$~(eV)&4.4&$4.2$ & $4.5$--$3.9$\,(Ga-dependent)\\ \hline
		$\Nc$~(cm$^{-3}$)&$3\times 10^{18}$ & $1.3\times 10^{18}$& $6.8\times 10^{17}$\\ \hline
		$\Nv$~(cm$^{-3}$) &$1.7\times 10^{19}$&$9.1\times 10^{19}$ &$1.5\times 10^{19}$\\ \hline
		$\ND$~(cm$^{-3}$)&$1\times10^{17}$\, (donor)  &$5\times10^{17}$\, (donor) &$2\times10^{16}$\, (acceptor)\\  \hline
		$\mun$~(cm$^2$ &$100$&$72$ &$100$\\ 
		V$^{-1}$~s$^{-1}$) && &\\ \hline
		$\mup$~(cm$^2$ &$31$&$20$ &$13$\\ 
		V$^{-1}$~s$^{-1}$) && &\\ \hline
		$\epsdc$ &$9$&$5.4$ & $13.6$\\ \hline
		$\Nf$~(cm$^{-3}$)& $10^{16}$& $5\times10^{17}$ & $10^{13}$--$10^{16}$ (Ga-dependent)\\ \hline
		$\sfE_{\rm T}$ &midgap& midgap & midgap\\ \hline
		$\sigman$~(cm$^2$)&$5\times10^{-13}$& $5\times10^{-13}$& $5\times10^{-13}$\\ \hline
		$\sigmap$~(cm$^2$)&$10^{-15}$& $10^{-15}$& $10^{-15}$\\ \hline
		$\RB$~(cm$^{3}$&$10^{-10}$ & $10^{-10}$&$10^{-10}$\\  
		s$^{-1}$)& & &\\  \hline
		{$\vth$~(cm~s$^{-1}$)} & {$10^7$} & {$10^7$} & {$10^7$}\\
		\hline \hline
	\end{tabular}
\end{table}

\subsection{Optoelectronic optimization} \label{sec:JVEF}	
Solution of the drift-diffusion equations enabled the calculation of the current density 
\begin{equation}
J(z) = \Jn(z) + \Jp(z)\, 
\end{equation}
flowing through the iZnO/CdS/CIGS {region}. Under steady-state conditions, $J(z)=\Jdev$ is constant throughout the solar cell. 
Thus, $\Jdev$ is the current density delivered to an external circuit;
$\Jsc$ is the value of $\Jdev$ when $\Vext=0$ and
$\Voc$ is the value of $\Vext$ such that $\Jdev=0$. With  the power density defined as
$P=\Jdev\Vext$, the maximum power density
$\Pmax$  obtainable from the solar cell is the highest point on the $P$-$\Vext$ curve.
The efficiency   is calculated as the ratio
${\Pmax}/{\Pin}$, where $\Pin = 1000$~W~m$^{-2}$   is the integral of  $S(\lambdao)$ over the solar spectrum. {Also,
a figure of merit called fill factor $FF={\Pmax}/{\Voc\Jsc}$ is commonly considered in solar-cell research.}
The DEA \cite{DEA} was used to optimize $\eta$, using a custom algorithm implemented with MATLAB\textsuperscript{\textregistered} version R2017b.

\section{Numerical results and discussion}\label{sec:OptoElecRes}

\subsection{Conventional CIGS solar cell (model validation)}\label{sec:Ref-conventional-cell}

First, we validated our coupled optoelectronic model by comparison
with extant experimental results for
the conventional MgF$_2$/AZO/iZnO/CdS/CIGS/Mo solar cell
containing a 2200-nm-thick homogeneous CIGS layer and a flat backreflector~\cite{abushama2005}. 
Values of  $\Jsc$, $\Voc$,   $FF$, and $\eta$ obtained from our model for
${\xi}=0$ ($\eg= 0.947$~eV), ${\xi}=0.25$ ($\eg= 1.12$~eV), and ${\xi}=1$ ($\eg= 1.626$~eV) are provided in Table~\ref{tab--ref-results}, as also are the corresponding experimental data \cite{ZWS, abushama2005}.  
The model predictions are in reasonable agreement with the experimental data,
the differences very likely due to variance between the optical and electrical properties used
in the model from those realized in practice.

\begin {table}[h]
\caption {\label{tab--ref-results}
 {Comparison of   $\Jsc$, $\Voc$,   FF, and $\eta$ predicted by
the coupled optoelectronic model 
for a conventional CIGS solar cell with a homogeneous CIGS layer (i.e., $A=0$)
with their experimental counterparts~\cite{abushama2005, ZWS}.}
}
\begin{center}
	\begin{tabular}{|p{0.4cm}|p{.6cm}|p{1.6cm}|p{0.8cm}|p{0.6cm}|p{0.4cm}|p{0.8cm}|}
		\hline \hline
	${\xi}$&  $\ego$ &  & $\Jsc$ &$\Voc$ & $FF$&$\eta$  \\
		&(eV)&  &(mA&(mV)&{(\%)}& {(\%)}\\
		&& &cm$^{-2})$&&& \\
		\hline 
0  &0.95  & \multirow{6}{*}{}Model      &38.63&497& 78& 15.05\\ 
 \cline{3-7}
&&Experiment     & & & & 	\\ 
&&~(Ref.~\cite {abushama2005})     &40.58&491&66&14.5	\\ 
\cline{3-7}
&&Experiment      & & & & 	\\ 

&&~(Ref.~\cite {abushama2005})  &41.1&491&75&15.0

\\ \hline

0.25&1.12 & \multirow{6}{*}{}Model  &34.41&648& 81& 18.12\\ 
\cline{3-7}
&&Experiment      & & & & 	\\ 
& &~(Ref.~\cite {abushama2005})  &35.22&692&79&19.5	\\ 
\cline{3-7}
&&Experiment      & & & & 	\\ 
&&~(Ref.~\cite {ZWS})   &37.8&741&81&22.6	\\  \hline

1 &1.626 & \multirow{6}{*}{}Model &14.86&911& 73& 9.92
\\ \cline{3-7}
&&Experiment      & & & & 	\\ 
& &~(Ref.~\cite {abushama2005})  &14.88&823&71&9.53	\\ 
\cline{3-7}
&&Experiment      & & & & 	\\ 
& &~(Ref.~\cite {abushama2005})  &18.61&905&75&10.2	\\
		\hline \hline
	\end{tabular}
\end{center}
\end {table}	

Next,  we assessed
the role of   traps at the  
CdS/CIGS interface in the solar cell \cite{Frisk14} by incorporating 
a 10-nm-thick surface-defect layer between the CdS and the CIGS layers
\cite{Songetal2004}. The interface trap density was taken to
be $10^{12}$~cm$^{-2}$, all other characteristics of the
surface-defect layer being the same as of the CIGS layer~\cite{Frisk14}.
The efficiency reduced in consequence, but the reduction was very
small. For example, the efficiency calculated for ${\xi}=0.25$ reduced from
18.12\% (Table~\ref{tab--ref-results}) to 18.11\%. This is accord with an 
experimental study concluding the  influence of surface recombination on high-efficiency CIGS solar cells to be insignificant~\cite{Kuciauskas2013}. We ignored the surface-defect layer for all results presented from hereonwards in this paper.

\subsection{Effect of $\Al2O3$ layer}\label{Al203_passivation}

To delineate the effect of the 50-nm-thick $\Al2O3$ layer between the CIGS layer
and a flat backreflector, we optimized the CIGS solar cell
with and without  that layer. Values of  $\Jsc$, $\Voc$,   $FF$, and $\eta$ obtained from our model 
for   $\LCIGS=400$~nm 
are presented in Table~\ref{tab--Al2O3--passivation}. 
The optimal efficiency is $12.49$\% with the $\Al2O3$ layer and $11.88$\% without it. 
 Surface recombination being insignificant~\cite{Kuciauskas2013}, the $5.1$\%-enhancement of $\eta$  is very likely due to the reduction of optical mismatch between 
Mo and CIGS by 
the $\Al2O3$ layer. Improvements in both short-circuit current density and open-circuit voltage due to the  $\Al2O3$ layer
can also be noted in Table~\ref{tab--Al2O3--passivation}. Hence, the 50-nm-thick $\Al2O3$ layer
is present in the solar cell for all results presented
 from now onward.

\begin {table}[htb]
\caption {\label{tab--Al2O3--passivation} 
 {Predicted parameters of the optimal CIGS solar cell
with and without the $\Al2O3$  layer when the 400-nm-thick CIGS layer is homogeneous
	($\ego\in[0.947, 1.626]$~eV and $A=0$) and the Mo backreflector is flat  ($\Lg=0$).}
}
\begin{center}
	\begin{tabular}{|p{0.8cm} |p{0.8cm}|p{1.cm}|p{0.8cm}| p{0.8cm}|p{0.8cm}|}
		\hline
		\hline
		${\La}$  &  $\ego$  & $\Jsc$ &$\Voc$ & $FF$&$\eta$  \\
		(nm) & (eV) &(mA &(mV)&{(\%)}& {(\%)}\\  [-18pt]
		& &\,\,cm$^{-2})$& & &  \\
		\hline
		 0   &1.27 &21.63&705& 77& 11.88\\ \hline
		50  &1.27 &22.65&711& 77& 12.49\\ 
		\hline
		\hline
	\end{tabular}
\end{center}
\end {table}


\subsection{Optimal solar cell: Homogeneous bandgap \& flat backreflector}\label{sec:opto_elechomo_FBR}
		
To allow comparison with a sensible baseline and highlight the advantages of the proposed designs, 
it is useful to run the optoelectronic optimization for a CIGS solar cell in which the bandgap is homogeneous
and the backreflector is flat; i.e., $A=0$  and $\Lg=0$. The parameter space
for  optimizing $\eta$ then reduces to: $\ego\in[0.947, 1.626]$~eV.  

Values of  $\Jsc$, $\Voc$,   $FF$, and $\eta$ \cite{Fonash,Jenny_Book} corresponding to the optimal design for
$\LCIGS\in\left\{100, 200, 300, 
400, 500,\right.$ $\left. 600, 900, 1200, 2200\ric$~nm are shown in Table~\ref{tab--homo-flat}. 
Depending on $\LCIGS$, the optimal homogeneous bandgap varies, with $\eg\in[1.24, 1.28]$~eV. 
The optimal efficiency increases with  $\LCIGS$. 
An efficiency of 13.79\% is predicted with an ultrathin-600-nm CIGS   layer.
The highest efficiency predicted is $18.93$\%, for a solar cell
with a 2200-nm-thick CIGS   layer
with an optimal bandgap of $\eg=1.24$~eV. 

		
\begin {table}[htb]
\caption {\label{tab--homo-flat} 
Predicted parameters of the optimal CIGS solar cell
with a specified value of 
$\LCIGS\in[100,2200]$~nm, when the CIGS layer is homogeneous
			($\ego\in[0.947, 1.626]$~eV and $A=0$) and the Mo backreflector is flat  ($\Lg=0$).
		}
	\begin{center}
			\begin{tabular}{|p{0.8cm}| p{0.8cm}|p{1.cm}|p{0.8cm}| p{0.8cm}|p{0.8cm}|}
				\hline
				\hline
				$\LCIGS$  &  $\ego$  & $\Jsc$ &$\Voc$ & $FF$&$\eta$  \\
				(nm) & (eV) &(mA &(mV)&{(\%)}& {(\%)}\\  [-18pt]
				& &\,\,cm$^{-2})$& & &  \\
				\hline
				100  &1.28 &14.89 &624 & 78 & 7.25\\\hline
				200  &1.26 &19.50 &660 & 76 & 9.76\\\hline
				300  &1.25&22.56 &681 & 76&11.59\\\hline
				400  &1.27 &22.65&711& 77& 12.49\\\hline
				500  &1.25 &23.71&704& 78& 13.15\\\hline
				600  &1.24 &24.66&704& 79& 13.79\\\hline
				900  &1.25 &25.68&725&80& 15.08\\\hline
				1200  &1.28 &25.72&756&81& 15.90\\\hline
				2200  &1.24 &31.11&742&82& 18.93\\
				\hline
				\hline
			\end{tabular}
	\end{center}
\end {table}
		

\begin {table}[htb]
\caption {\label{tab--homo-period} 
Predicted parameters of the optimal CIGS solar cell
with
 a specified value of 
$\LCIGS\in[100,2200]$~nm, when the CIGS layer is homogeneous
			($\ego\in[0.947, 1.626]$~eV and $A=0$) and the Mo backreflector is periodically corrugated.}
\begin{center}
	\begin{tabular}{ |p{0.6cm}| p{0.6cm}|p{0.6cm}|p{0.5cm}|p{0.6cm}|p{0.6cm}|p{0.4cm}|p{0.4cm}|p{0.5cm}|}
		\hline
		\hline
		$\LCIGS$ & $\ego$ & $\Lx$ &$\zeta$ & $\Lg$  & $\Jsc$ & $\Voc$&$FF$&$\eta$  \\
		(nm) & (eV) & (nm)& & (nm)&(mA& (mV)&{(\%)}&{(\%)} \\[-8pt]
		&  & & & &cm$^{-2}$)&&& \\
		\hline
		100 & 1.28 &500 &{0.50} &97 &14.89 &624 & 78 & 7.25\\ \hline
		200 & 1.26 &510 &0.50 &101&19.53 &661 & 76 & 9.91\\ \hline
		300 & 1.25   &510 &0.50 &101 &22.56&681 & 76&11.59\\ \hline
        400 & 1.27 &510 &0.49 &101 &22.79&711& 77& 12.58\\ \hline
		500 & 1.25 &510 &0.48 &106 &23.78&705& 78& 13.19\\ \hline
		600 & 1.24 &510 &0.48 &105& 24.69&704& 79& 13.81\\ \hline
		900  &1.25 &502 &0.49 &101&25.71&725&80& 15.09\\ \hline
		1200  &1.28& 502 &0.49 &101&25.72&759&81& 15.90\\ \hline
		2200  &1.24 &502 &0.49 &101&31.11&742&82& 18.93\\     
		\hline
		\hline
	\end{tabular}
\end{center}
\end {table}
\subsection{Optimal solar cell: Homogeneous bandgap \& periodically corrugated backreflector}\label{sec:opto_elechomo_PCBR}

Next,  we repeated optoelectronic optimization for solar cells containing a homogeneous CIGS layer but with a periodically corrugated backreflector instead of a flat one. 
The results of this optimization exercise  for fixed $\LCIGS$ are provided in Table~\ref{tab--homo-period}.
On comparing Tables~\ref{tab--homo-flat} and \ref{tab--homo-period}, we see that periodic corrugation of
the Mo backreflector improves the  efficiency  by no more than $2$\% (at $\LCIGS=200$~nm). This indicates the moderate benefit of exciting both surface-plasmon-polariton (SPP) waves \cite{Anderson1983,Heine1995} and waveguide modes \cite{Khaleque2013} by taking advantage of the grating-coupled configuration \cite{ESW2013}, when the CIGS layer is ultrathin. However, that benefit vanishes for thicker CIGS layers.

\subsection{Optimal solar cell: Linearly nonhomogeneous bandgap} \label{sec:Linearly_nonhomo}
Next, let us consider the maximization of $\eta$ as a function of $\LCIGS$ when the CIGS   layer has a
linearly  nonhomogeneous bandgap, according to either Eq.~\r{Eqn:Linear-bandgap} or Eq.~\r{Eqn:Linear-bandgap1}. 

\subsubsection{Forward  grading}
Equation~\r{Eqn:Linear-bandgap} is used for  
linearly nonhomogeneous forward bandgap  grading, so that
$\eg(\Lw+\LZnO+\LCdS)\leq\eg(\Ld)$ for $A\geq0$,  the bandgap being smaller near the front contact than near the back contact. 
Optoelectronic optimization yielded $A=0$. Thus, for forward   bandgap grading of the CIGS layer,
Table~\ref{tab--homo-flat} holds when the backreflector is flat and Table~\ref{tab--homo-period}
holds  when the backreflector is periodically corrugated.

\subsubsection{Backward  grading}
When  Eq.~\r{Eqn:Linear-bandgap} is replaced by Eq.~\r{Eqn:Linear-bandgap1} so that
$\eg(\Lw+\LZnO+\LCdS)\geq\eg(\Ld)$ for $A\geq0$, optoelectronic optimization
predicts $A > 0$ for optimal efficiency.

Table~\ref{tab--linear-grading} presents optimal $\eta$ for nine values of
$\LCIGS$, when the CIGS bandgap is  linearly nonhomogeneous
according to Eq.~\r{Eqn:Linear-bandgap1} and the Mo backreflector is periodically corrugated. 
For $\LCIGS=100$~nm, the optimal $\eta=9.88$\% in Table~\ref{tab--linear-grading},
whereas $\eta = 7.25$\% in Table~\ref{tab--homo-flat}.  This relative enhancement  of $36.27$\% must be attributed
to the   backward bandgap grading of the CIGS layer.  
Concurrently, $\Jsc$ increases from $14.89$~mW~cm$^{-2}$ to $15.09$~{mA}~cm$^{-2}$ ($1.3$\% relative increase) and 
$\Voc$  from $624$~mV to $960$~mV ($53.84$\% relative increase); however, the
fill factor reduces from $78$\% to $68$\%. 

 {The overall trend encompasses the enhancement of both  $\eta$ and $\Voc$
and the reduction of  $\Jsc$ and FF with  backward bandgap grading for all considered thicknesses of the CIGS layer.}
For $\LCIGS=600$~nm, upon the incorporation of a linearly nonhomogeneous  bandgap and a
periodically corrugated backreflector, the optimal efficiency increases from  $13.79$\% (Table~\ref{tab--homo-flat})
by $14.5$\% to  $15.79$\% (Table~\ref{tab--linear-grading}) and $\Voc$  increases from $704$~mV by {$45.31$\%} to {$1023$}~mV,
but $\Jsc$ decreases from $24.66$~mA~cm$^{-2}$ by $12.89$\% to $21.48$~mA~cm$^{-2}$ and the
fill factor reduces from $79$\% to $71$\%. 
The relative enhancement in efficiency decreases as $\LCIGS$ increases. Thus, the relative enhancement  is only $1.7$\% for $\LCIGS=2200$~nm. Higher values of $\Voc$ are positively correlated with   larger values of $\eg(\Lw+\LZnO+\LCdS)$.

 	\begin {table}[htb]
	\caption {\label{tab--linear-grading} 
Predicted parameters of the optimal CIGS solar cell
with a specified value of 
$\LCIGS\in[100,2200]$~nm,  when the CIGS layer is linearly nonhomogeneous
		according to Eq.~\r{Eqn:Linear-bandgap1} and the Mo backreflector is periodically corrugated.  }
	\begin{center}
		\begin{tabular}{ |p{1.1cm}| p{1cm}| p{1.2cm}|p{0.8cm}|p{0.8cm}|p{0.8cm}|p{0.8cm}|p{1.8cm}|p{1.2cm}|p{1cm}|p{1cm}|}
			\hline
			\hline
			$\LCIGS$ & $\egmax$ &$\ego$ & A & $\Lx$ &$\zeta$ & $\Lg$  &
			$\Jsc$ & $\Voc$&$FF$&$\eta$  \\
			(nm) & (eV)& (eV)& & (nm) & & (nm)&(mA~cm$^{-2}$)& (mV)& {(\%)}& {(\%)} \\
			\hline
			100 & 1.61 &0.96 &0.98 &500 &0.50 &97& 15.09 &960 &  68 &9.88\\ \hline
			200 & 1.61 &0.96 &0.98 &510 &0.51 &101& 19.28 &995 & 62&12.08\\ \hline
			300 & 1.62 &0.96 &0.98 &502 &0.49 &101& 20.70 &1010 & 63&13.34\\ \hline
			400 & 1.62 &0.96 &0.98 &510 &0.49 &101& 21.13 &1011 &  67&14.34\\   \hline 
			500 & 1.62 &0.95 &0.99 &510 &0.48 &106& 21.31 &1017&  69&15.14\\   	 \hline
			600 & 1.62 &0.95 &0.75 &500 &{0.50}&101& 21.48 &1023 &  71&15.79\\   \hline
			900 & 1.62 &0.95&0.75 &500 &{0.50} &101& 22.21 &1032 &  75&17.24\\   \hline
			1200 & 1.62 &0.95 &0.75 &500 &{0.50}&101& 22.74 &1037 &  76&18.07\\  \hline
			2200 & 1.62 &0.95 &0.75 &500 &{0.50} &101& 24.09 &1039 &  77&19.27\\  			
			\hline
			\hline
		\end{tabular}
	\end{center}
	\end {table}
 

\subsection{Optimal solar cell: Sinusoidally nonhomogeneous bandgap \& periodically corrugated backreflector}\label{sec:optoelec_nonhomo_PCBR}
Next, let us consider the optoelectronic optimization of  $\eta$  for fixed values of $\LCIGS$ for a solar cell with a periodically nonhomogeneous CIGS layer according to Eq.~\r{Eqn:Sin-bandgap} and a periodically corrugated backreflector. Values of  $\eta$ optimized for fixed $\LCIGS$ are shown in Table~\ref{tab--sinu-period}.
Values of $\ego$, $A$, $\alpha$, $K$, $\psi$, $\Lx$, $\zeta$, $\Lg$,   $\Jsc$, $\Voc$, and $FF$ for the
optimal designs are also shown in this table.

For $\LCIGS=100$~nm, the optimal efficiency is $12.37$\% in Table~\ref{tab--sinu-period},
a relative increase of $70.62$\% over the optimal efficiency of $7.25$\%  in Table~\ref{tab--homo-flat}.
This enhancement must be due to the sinusoidally nonhomogeneous bandgap.  Concurrently,
$\Jsc$ increases from $14.89$~mA~cm$^{-2}$ to $17.04$~mA~cm$^{-2}$ ($14.43$\% relative increase) and 
$\Voc$  from $624$~mV to $969$~mV ($55.28$\% relative increase); however, the
fill factor reduces from $78$\% to $71$\%. 
 {Notably, $\Voc$ increased significantly without   $\Jsc$ decreasing, which is in contrast to solar cells with either a homogeneous or a linearly nonhomogeneous CIGS layer~\cite{Gloeckler-Sites2005, Songetal2004}. }
 
For $\LCIGS=200$~nm, upon the incorporation of a sinusoidally nonhomogeneous  bandgap and a
periodically corrugated backreflector, the optimal efficiency increases from $9.76$\% (Table~\ref{tab--homo-flat}) by $73.05$\% to $16.89$\% (Table~\ref{tab--sinu-period}), $\Jsc$ increases $19.50$~mA~cm$^{-2}$ by
$23.69$\% to $24.12$~mA~cm$^{-2}$, and $\Voc$  from $660$~mV by $52.57$\% to $1007$~mV, but the
fill factor reduces from $76$\% to $69$\%. 
Likewise, for $\LCIGS=600$~nm, the optimal efficiency increases  by $65.98$\% to {$22.89$\% from $13.79\%$}, $\Jsc$ increases by $18.32$\% to $29.18$~mA~cm$^{-2}$ from {$24.66$}~mA~cm$^{-2}$, and $\Voc$ increases by $48.43$\% to $1045$~mV from $704$~mV, although the fill factor reduces to $75$\% from $79$\%. The overall trend  is that the relative enhancement of $\eta$---due 
to the sinusoidally nonhomogeneous bandgap and the periodically corrugated 
banckreflector---decreases with the increase of  $\LCIGS$, on optoelectronic optimization.    
The highest optimal $\eta= 27.7$\% in Table~\ref{tab--sinu-period} was obtained with the conventional 2200-nm-thick CIGS   layer, a relative enhancement of $46.32$\% with respect to $\eta=18.93$\% for
the   homogeneous CIGS layer in Table~\ref{tab--homo-flat}.   The short-circuit current density increases from $31.11$~mA~cm$^{-2}$ by $6.5$\% to $33.16$~mA~cm$^{-2}$, $\Voc$ increases from $742$~mV by $44.2$\% to $1070$~mV, but the fill factor reduces to $78$\% from $82$\%.

Optimal values of $A$ range from $0.98$ to $1$ in Table~\ref{tab--sinu-period}, which is in contrast to $A<0.04$ delivered by optical optimization of $\Psup$ \cite{Ahmad2018}.  
The maximum value of $A$   provides the largest possible bandgap variation in the nonhomogeneous CIGS layer. 
Thus, optoelectronic optimization delivers sinusoidal nonhomogeneity
of the bandgap with a large amplitude, whereas optical optimization severely suppresses
nonhomogeneity of the bandgap.  The inescapable
conclusions are that   (i) optical optimization is seriously deficient and (ii) optoelectronic optimization
is essential for nonhomogeneous-bandgap solar cells.  	

Previous studies on Schottky-barrier thin-film solar cells~\cite{Anderson2017, Anderson2018} had suggested that optimal values of $K$ are integer multiples of $1.5$ and
that $\psi=0.75$. Both
predictions are mostly upheld by the optimal data in Table~\ref{tab--sinu-period}. 

 The relative enhancement in $\Voc$ is almost the same for   linear bandgap grading (Table~\ref{tab--linear-grading}) as for sinusoidal bandgap grading (Table~\ref{tab--sinu-period}) of the CIGS layer 
in comparison to the homogenous CIGS layer (Tables~\ref{tab--homo-flat} and \ref{tab--homo-period}).
However, whereas sinusoidal bandgap grading of the CIGS layer enhances $\Jsc$, linear bandgap grading of that layer
actually depresses $\Jsc$, in comparison to the homogeneous CIGS layer. As the reduction of $\Jsc$
does not completely overcome the enhancement of $\Voc$ for linear bandgap grading,
the efficiencies in Table~\ref{tab--linear-grading} exceed their counterparts
in Tables~\ref{tab--homo-flat} and \ref{tab--homo-period}; of course, the efficiencies
in Table~\ref{tab--sinu-period} are even higher.  We conclude that sinusoidally nonhomogeneous bandgap
is more efficient than the homogeneous and the linearly nonhomogeneous bandgaps for all  $\LCIGS\in[100,2200]$~nm.

 
	\begin {table}[htb]
	\caption {\label{tab--sinu-period} 
	Predicted parameters of the optimal CIGS solar cell
 with a specified value of 
$\LCIGS\in[100,2200]$~nm, 
when the CIGS layer is sinusoidally nonhomogeneous
		according to Eq.~\r{Eqn:Sin-bandgap} and the Mo backreflector is periodically corrugated. 
	}
	\begin{center}
		\begin{tabular}{ |p{1.1cm}| p{0.7cm} |p{0.8cm}| p{0.8cm}| p{0.8cm}|p{0.8cm}|p{0.8cm}|p{0.8cm}|p{0.8cm}|p{1cm}|p{1cm}|p{1cm}|p{1cm}|}
			\hline
			\hline
			$\LCIGS$ & $\ego$  & A &$\alpha$& $K$ & $\psi$ & $\Lx$ &$\zeta$ & $\Lg$  &
			$\Jsc$ & $\Voc$&  $FF$ &$\eta$  \\
			(nm) & (eV) & & & & & (nm)& & (nm)&(mA& (mV)&{(\%)}& {(\%)}\\
			&   & & & & &  & &  & cm$^{-2}$)& & &  \\
			\hline
			100 & 0.96 & 0.99 &6.14 & 0.75 & 0.75 &510 &{0.50} &101& 17.04 &969  &71 &12.37\\ \hline
			200 & 0.95 & 1.00 & {6.0} & {1.50} & 0.75 &500 &0.48 &101& 24.12 &1007  &69 &16.89 \\	\hline			
			300 & 0.95 & 0.98 & {6.0} & {1.50} & 0.74 &502 &0.49 &101& 25.98 &1023 &71 & 19.01\\ \hline
			400 & 0.95 & 0.98 & {6.0} & {1.50} & 0.75 &500 &{0.50} &111& 27.17 &1033 &  73 & 20.66\\  \hline
			500 & 0.95 & 0.99 & {6.0} & {1.50} & 0.76 &520 &{0.50} &111& 28.23 &1040  &74 & 21.90\\ \hline
			600 & 0.95 & {0.99} & {6.0} & {1.50} & 0.75 &510 &0.48 &106& 29.18&1045&75 & 22.89\\  \hline
			900 & 0.95 & {0.99} & {6.0} & {1.50} & 0.75 &510 &0.48 &106& 30.86&1057&76 &24.98\\  \hline
			1200 & 0.95 & 0.98 & {6.0} & {1.50} & 0.75 &510 &0.48 &106& 32.02&1063&77 &26.33\\ \hline	
			2200 & 0.95 & 0.98 & {6.0} & {1.50} & 0.75 &510 &0.48 &106& 33.16&1070&78 & {27.70}\\  
			\hline
			\hline
		\end{tabular}
	\end{center}
	\end {table}
 

\subsection{Optimal solar cells with 600-nm-ultrathin CIGS layer} \label{sec:Optimal_design}

 {The highest efficiency in Tables~\ref{tab--homo-flat}--\ref{tab--sinu-period} is 27.7\%. It is
predicted by our coupled optoelectronic model for  
the  
CIGS solar cell with a sinusoidally nonhomogeneous 2200-nm-thick CIGS layer. However, we are interested in ultrathin CIGS  
layers to reduce the material and processing costs, keeping in mind the scarcity of indium.
The solar cell with the sinusoidally nonhomogeneous CIGS layer of 600-nm thickness and a periodically
corrugated Mo backreflector has efficiency $\eta=22.89$ which   compares well with the efficiency
of the conventional CIGS solar cell  with a 2200-nm-thick homogeneous   CIGS layer~\cite{ZWS}. 
Therefore, a detailed study of the solar cell with
the 600-nm-thick   nonhomogeneous   CIGS   layer   is reported next.}

\subsubsection{{Backward} linearly nonhomogeneous bandgap}

The design and performance parameters of the optimal 
{CIGS} solar cell with a {600}-nm-thick linearly nonhomogeneous
CIGS layer are provided in Table~\ref{tab--linear-grading}. Spatial profiles of $\eg(z)$
and $\chi(z)$ delivered
by optoelectronic optimization are provided in Fig.~\ref{back-Eg}(a).

Figure~\ref{figure-EcEvEinpni-linNonH}(a) presents the spatial profiles of  
$\Ec(z)$, $\Ev(z)$, and $\Ei(z)$.  The spatial variations of $\Ec$ and $\Ei$ are similar to that
of $\eg$ and provide the conditions to enhance the generation rate \cite{Gloeckler-Sites2005}.
Figure~\ref{figure-EcEvEinpni-linNonH}(b) presents the graphs of $n(z)$, $p(z)$, and $\ni(z)$
under the equilibrium condition. 
The intrinsic carrier density varies linearly such that it is small  where $\eg$ is large and \textit{vice versa}.

%
\begin{figure}[htb] 
	\centering   
	\includegraphics[width=0.4\textwidth]{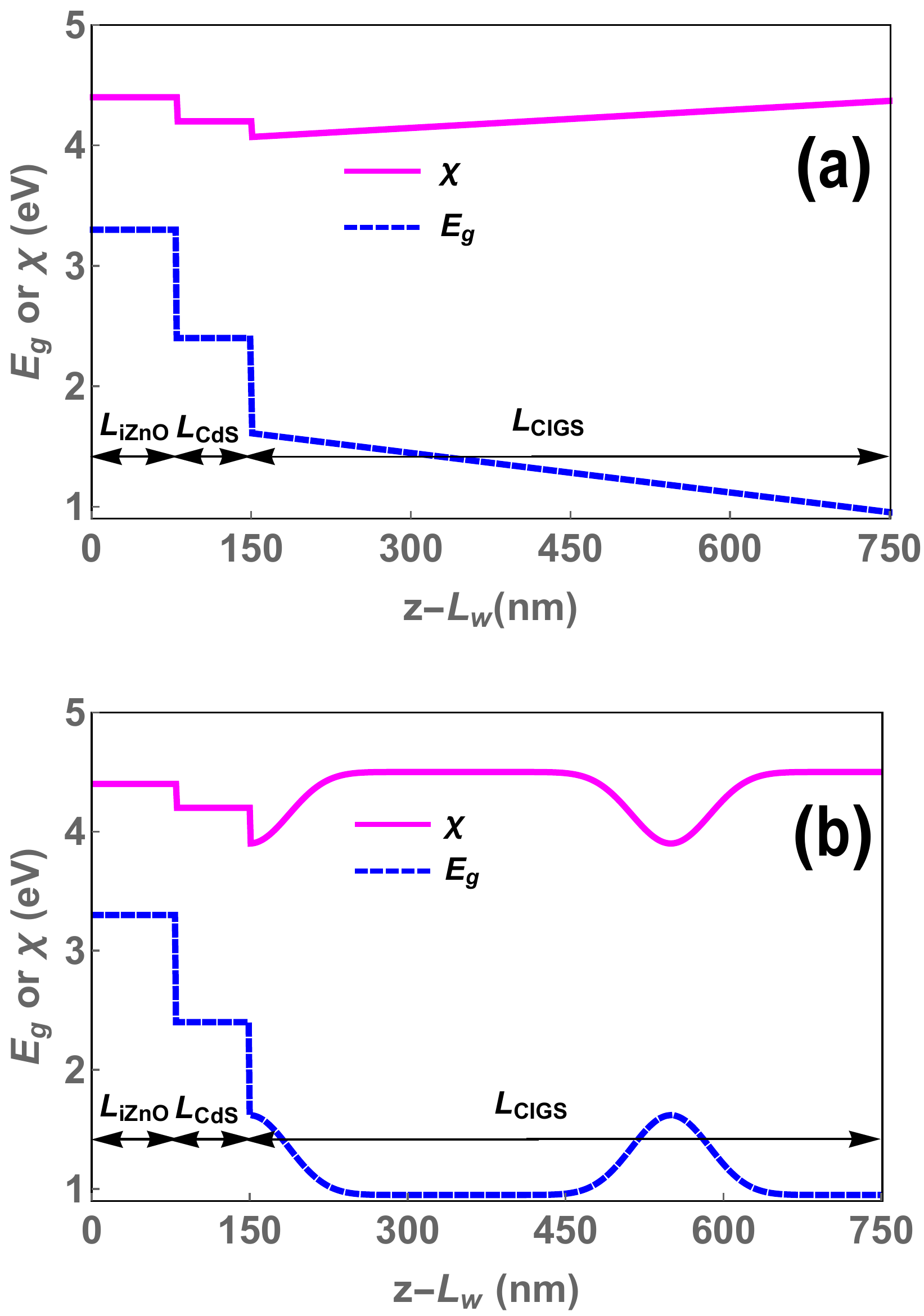} 
	\caption{ Variation of $\eg$ and $\chi$ with $z$ in the iZnO/CdS/CIGS region of the optimal CIGS solar cell with a 600-nm-thick (a) linearly nonhomogeneous or (b)  sinusoidally nonhomogeneous CIGS layer.  The design parameters   are available in Tables~\ref{tab--linear-grading} and ~\ref{tab--sinu-period}.
		\label{back-Eg} }
\end{figure}
%

Spatial profiles of 
$G(z)$ and $\Rnpz$ are given in Fig.~\ref{back-GRJV}(a).
The  generation rate is higher near the front face $z=\Lw+\LZnO+\LCdS$ and  the back face
$z=\Ld$ of the CIGS layer and slightly lower in the middle of that layer, but
the recombination rate drops sharply near the back face of that layer.
Furthermore, the spatial profile of the recombination rate follows that of the 
{defect} density $\Nf$ (not provided here). 
The $\Jdev$-$\Vext$ characteristics of the solar cell are shown in Fig.~\ref{back-GRJV}(b). 
From this figure, ${\Jdev}=17.45$~mA~cm$^{-2}$, ${\Vext}=0.905$~V,  
$FF=73$\%,  and $\eta=15.79$\% {for the best performance}.

\begin{figure}[htb] 
	\centering   
	\includegraphics[width=0.4\textwidth]{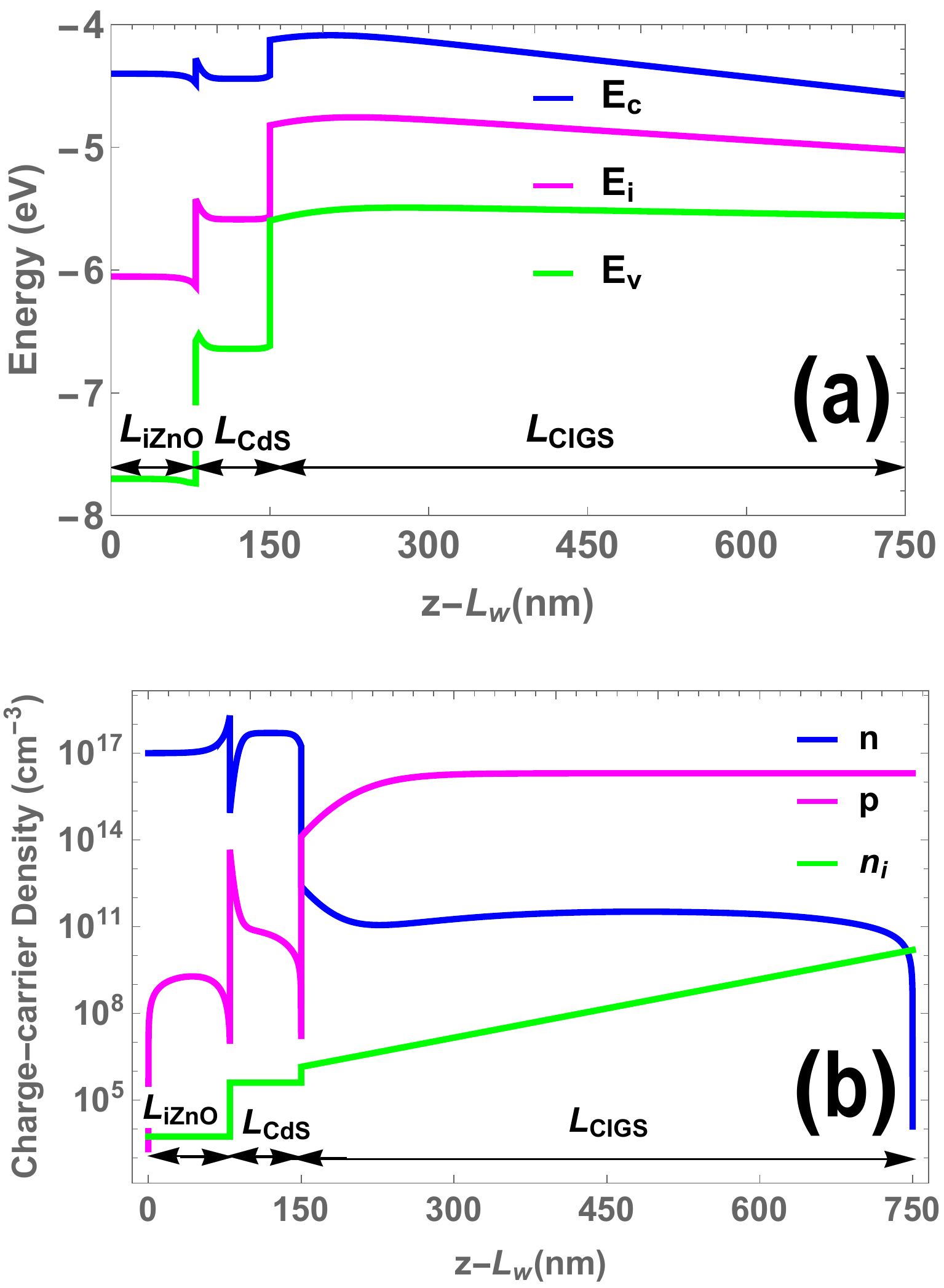} 
	\caption{(a) Spatial profiles of $\Ec(z)$, $\Ev(z)$, and $\Ei(z)$ in the iZnO/CdS/CIGS 
	{region} of the 
		optimized 600-nm-thick linearly graded bandgap CIGS solar cell (b) Spatial profiles of $n(z)$, $p(z)$, 
		and $\ni(z)$ in the iZnO/CdS/CIGS {region} of the optimized 600-nm-thick linearly graded bandgap CIGS solar cell. 
		\label{figure-EcEvEinpni-linNonH} }
\end{figure}	
%
\begin{figure}[htb] 
	\centering   
	\includegraphics[width=0.4\textwidth]{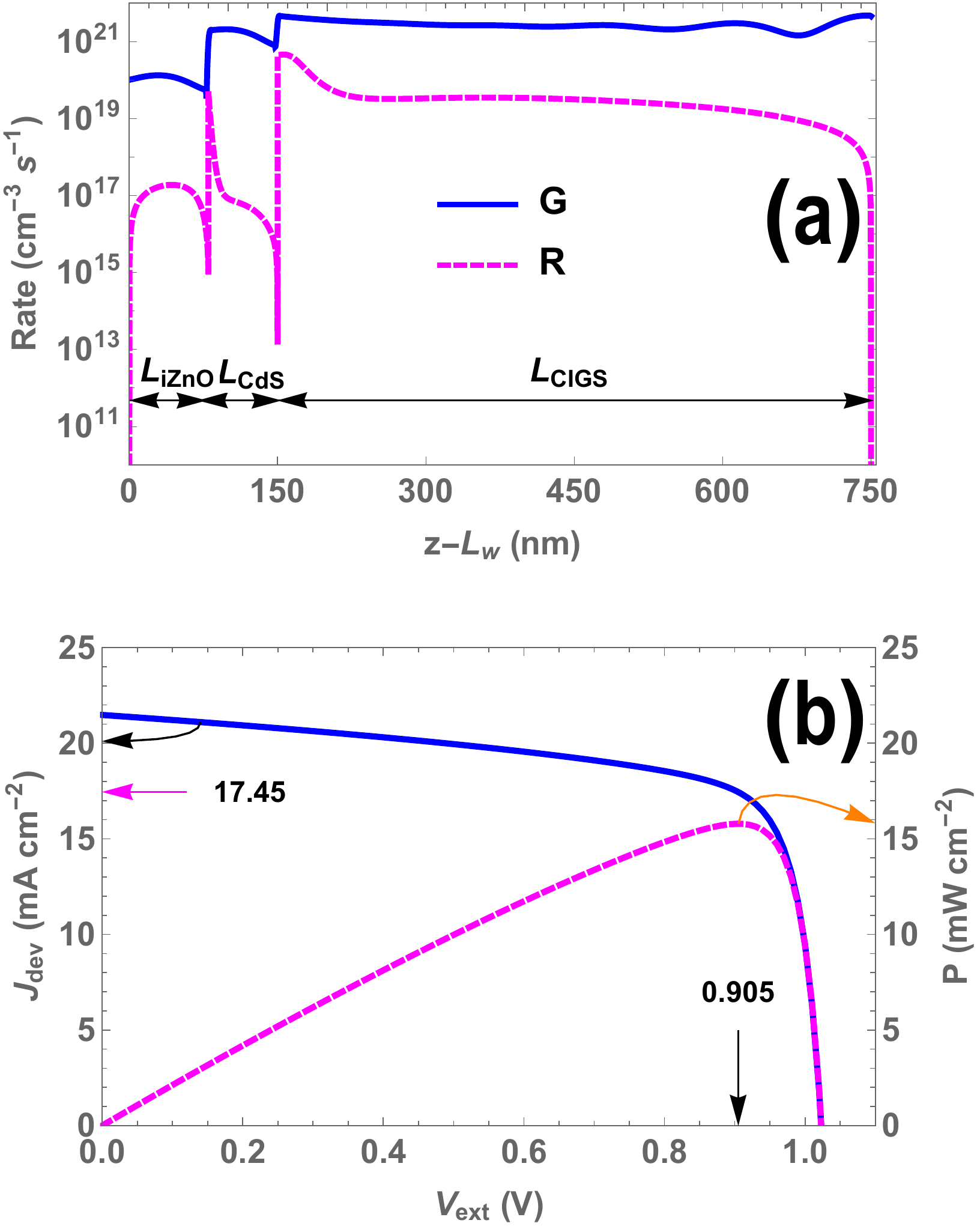} 
	\caption{ (a) Spatial profiles of   $G(z)$ and   $\Rnpz$ 
		in the iZnO/CdS/CIGS region of the optimal CIGS solar cell with a 600-nm-thick linearly nonhomogeneous
		CIGS layer. 
		The design parameters of this solar cell are available in Table~\ref{tab--linear-grading}. 
		(b)  Plots of $\Jdev$ and $P$ vs. $\Vext$ of the solar cell. The blue arrow identifies the $\Jdev$-$\Vext$ curve
		and the orange arrow identifies the $P$-$\Vext$ {curve}. The numerical values of $\Jdev$ and $\Vext$ for maximum $P$ 
		are also identified. 
		\label{back-GRJV} }
\end{figure}
%

\subsubsection{Sinusoidally nonhomogeneous bandgap \& periodically corrugated backreflector}\label{sec3.G.2}
 {The parameters of the  optimal CIGS solar cell with a 600-nm-thick sinusoidally nonhomogeneous
CIGS layer are provided in Table~\ref{tab--sinu-period}.} The variations of  $\eg$ and $\chi$  with $z$ in 
the  iZnO/CdS/CIGS region of the ultrathin CIGS
solar cell are depicted in Fig.~\ref{back-Eg}(b). With $\ego=0.95$~eV and $A=0.992$, $\eg(z)\in[0.95,1.62]$~eV. 
The magnitude of $\eg(z)$ is large in  the proximity of the
plane $z=\Lw+\LZnO+\LCdS$, which elevates $\Voc$. The regions
in which $\eg(z)$ is  small are of substantial thickness, these regions being
responsible for elevating the electron-hole-pair generation rate~\cite{Fonash}. 
The sinusoidal grading close to the back-surface adds additional drift field to reduce 
the back-surface recombination rate, supplementing the role of the $\Al2O3$ passivation layer.
Thus the bandgap profile is ideal  for high efficiency. 	

Figure~\ref{figure-EcEvEinpni-SinNonH}(a) shows the variations of $\Ec$, $\Ev$, and $\Ei$ with respect to $z$.
The spatial profiles of $\Ec$ and $\Ei$ are similar to that of $\eg$ and provide the conditions to enhance the generation rate \cite{Gloeckler-Sites2005}.		
Figure~\ref{figure-EcEvEinpni-SinNonH}(b) shows the spatial variations
of the electron, hole and intrinsic carrier densities under equilibrium condition. 
The holes are the majority carriers in {$p$}-type CIGS and electrons are the majority carriers in {$n$}-type CdS. 
The intrinsic carrier density  varies sinusoidally such that
$\ni$ is small  where $\eg$ is large and \textit{vice versa}.

\begin{figure}[htb] 
	\centering   
	\includegraphics[width=0.4\textwidth]{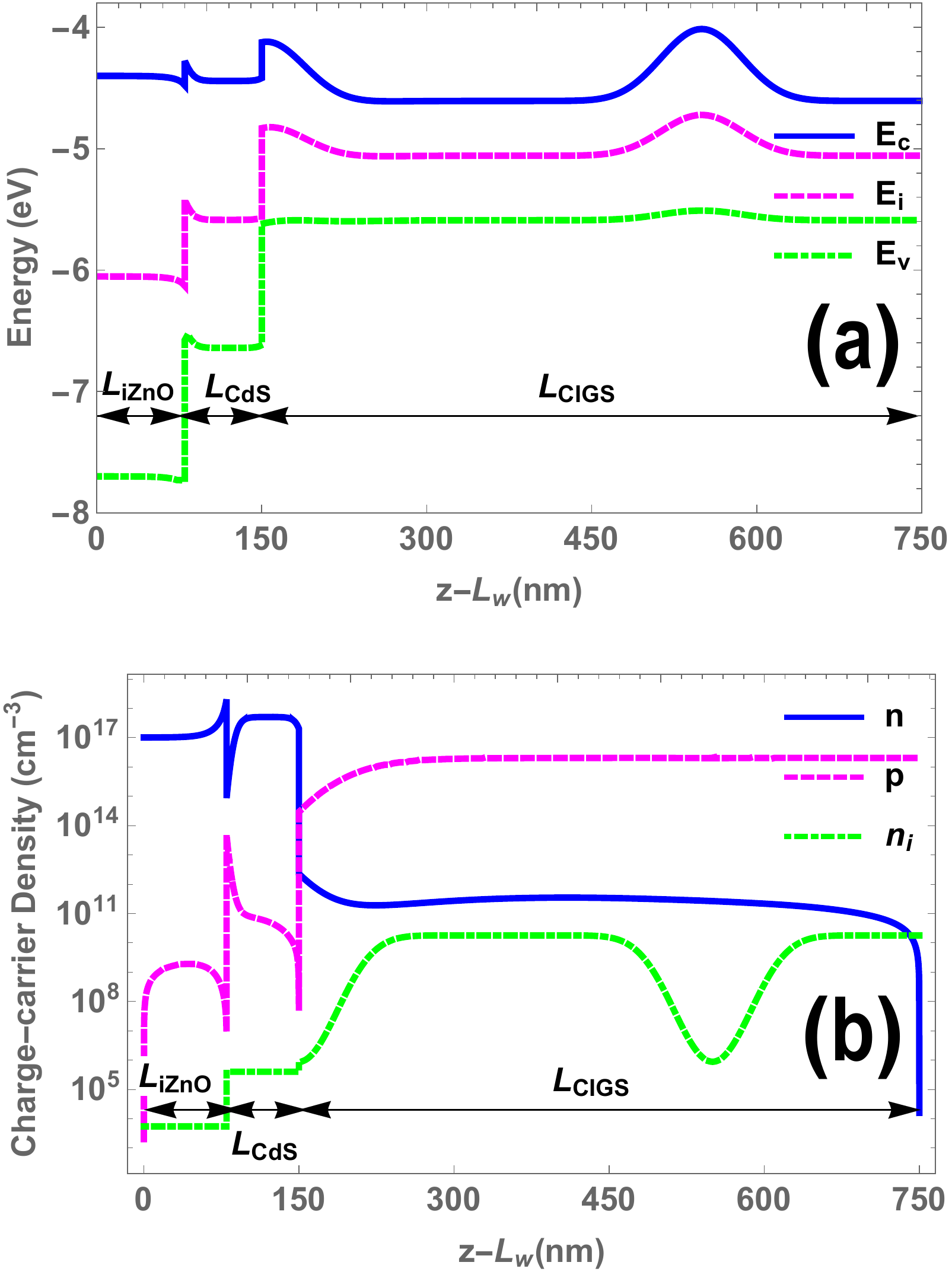} 
	\caption{Same as Fig.~\ref{figure-EcEvEinpni-linNonH} except that the bandgap is
		varying sinusoidally according to Eq.~(\ref{Eqn:Sin-bandgap}). The design parameters of this solar cell are available in Table~\ref{tab--sinu-period}.
		\label{figure-EcEvEinpni-SinNonH} }
\end{figure}

Profiles of the electron-hole-pair generation rate $G(z)$ and recombination rate
$\Rnpz$ are shown in Fig.~\ref{sinu-GRJV}(a). The generation rate is higher in regions with
lower bandgap and \textit{vice versa}.
Higher electron-hole-pair generation rate in the proximity of the plane $z=\Ld$ can be seen as a consequence of 
the enhanced   electric field  at optical frequencies
due to the excitation of  SPP waves near the back face of the CIGS layer facilitated by the periodically corrugated backreflector {\cite{Anderson1983,Heine1995,Shuba2,Shoji_paper,Liu2015}.}  	
	
The $\Jdev$-$\Vext$   characteristics of the solar cell are shown in Fig.~\ref{sinu-GRJV}(b).  Our optoelectronic model predicts $\Jdev=24.72$~mA~cm$^{-2}$, $\Vext=0.926$~V, $FF=76$\%, and   $\eta=22.89$\% for {the best performance of} this solar cell. Given that the   thickness of the CIGS layer is just $600$~nm, this value of   efficiency compares favorably with the $22$\% efficiency reported for solar cells with  a
$2200$-nm-thick homogeneous CIGS layer \cite{ZWS,Green2018}.

\begin{figure}[htb] 
	\centering   
	\includegraphics[width=0.4\textwidth]{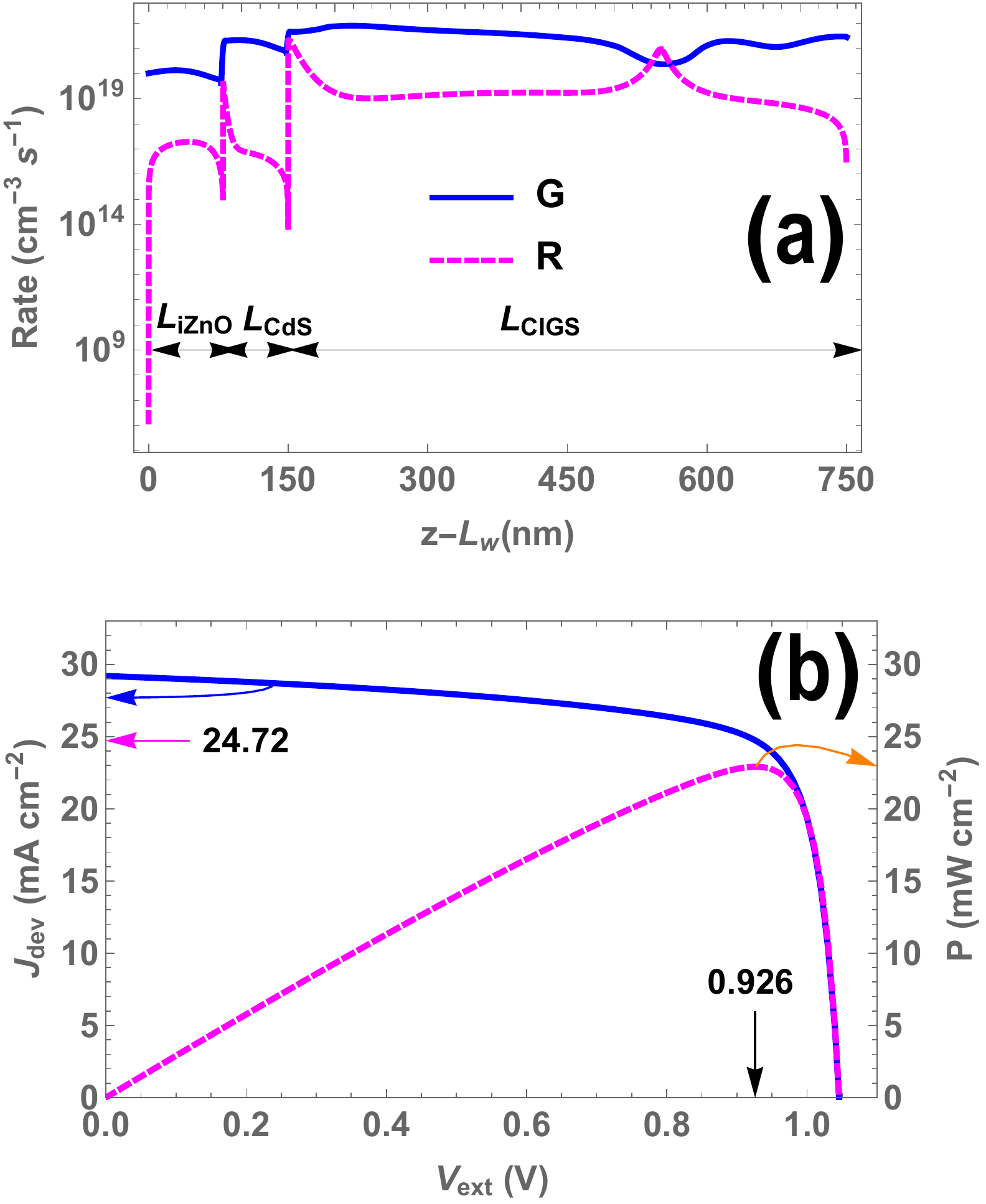} 
	\caption{Same as Fig.~\ref{back-GRJV} except that the bandgap is varying sinusoidally according to Eq.~(\ref{Eqn:Sin-bandgap}). 	
		The design parameters of this solar cell are available in Table~\ref{tab--sinu-period}.
		\label{sinu-GRJV}	 }
\end{figure}

\subsubsection{Sinusoidally nonhomogeneous bandgap vs. linearly nonhomogeneous bandgap}

When the bandgap is sinusoidally graded in the CIGS layer, the electron-hole-pair generation rate is higher in the  
small-bandgap regions  than elsewhere in the CIGS layer.  The open-circuit voltage is elevated in the optoelectronically
optimal designs, because the bandgap is high in  the proximity of both faces of the CIGS layer~\cite{Anderson2018,Gloeckler-Sites2005,Song2010}. Both of these features help increase the efficiency,  as discussed in Sec.~3.\ref{sec3.G.2}.

Similar conclusions emerge when the bandgap is linearly graded in the
CIGS layer, with the bandgap being smaller near the back face than near the front face.  However,
Fig.~\ref{back-Eg}(b) shows that the bandgap is flat and low ($\sim$1~eV) in a large portion of the CIGS layer
when the bandgap is sinusoidally nonhomogenous, but that feature is missing
in Fig.~\ref{back-Eg}(a) for the linearly nonhomogeneous CIGS layer. The spatial profiles of   $G(z)-\Rnpz$ provided in Fig.~\ref{GminusR} confirm that the net production of charge carriers is boosted in regions of uniformly low bandgap. Hence, the efficiency is
  significantly higher   for   sinusoidal grading ($22.89\%$)  than for linear grading ($15.79\%$)
of the bandgap in the CIGS layer.

\begin{figure}[htb] 
	\centering   
	\includegraphics[width=0.4\textwidth]{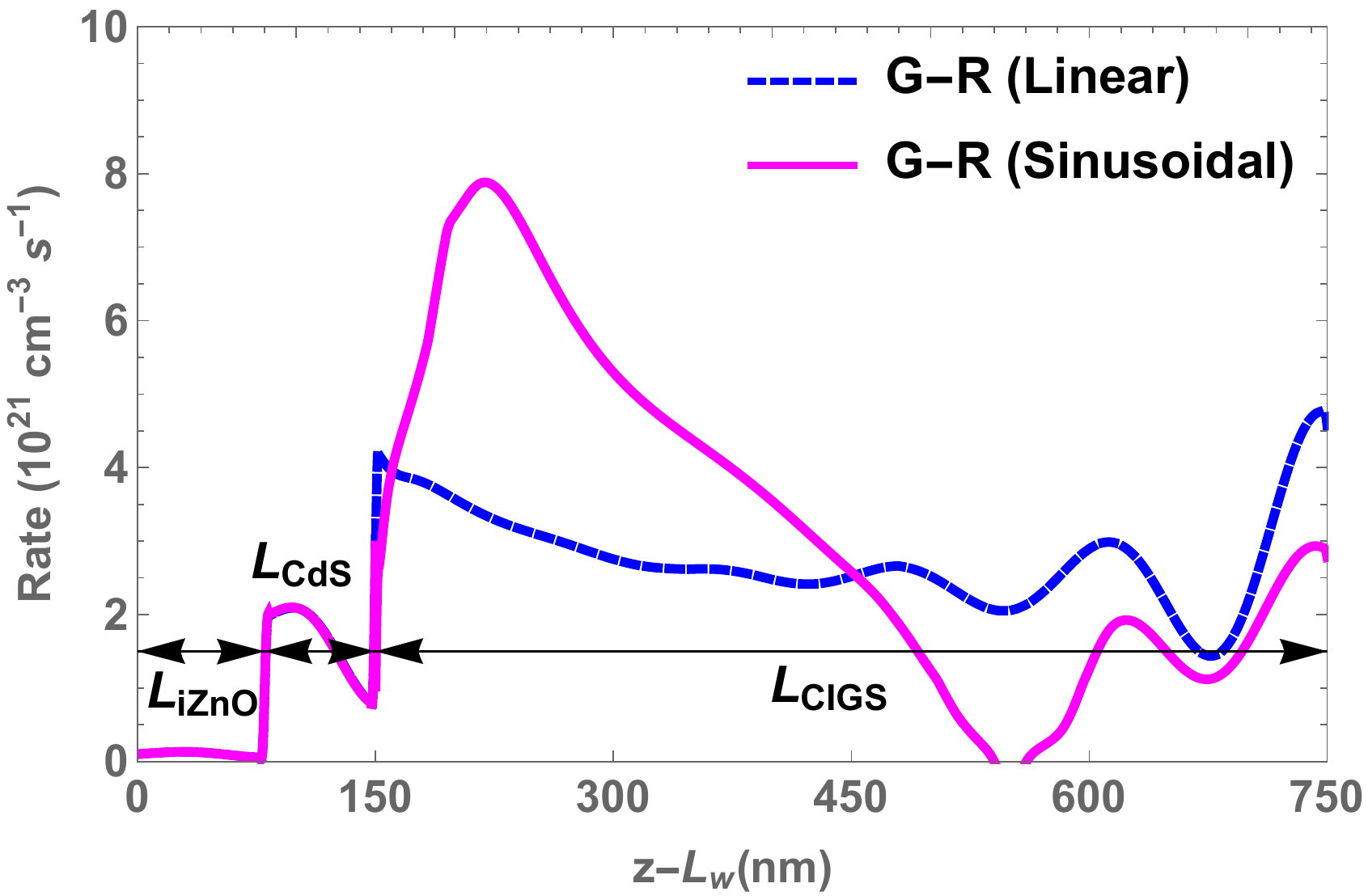} 
	\caption{Spatial profiles of   $G(z)-\Rnpz$ 
		in the iZnO/CdS/CIGS region of the optimal CIGS solar cell with a 600-nm-thick 		CIGS layer in which the bandgap is either linearly (Table~\ref{tab--linear-grading})
		or sinusoidally (Table~\ref{tab--sinu-period}) graded.
		\label{GminusR} }
\end{figure}
%

\section{Concluding remarks}\label{sec:conc}
	
Optoelectronic optimization was carried out for an  ultrathin  CuIn$_{1-\xi}$Ga$_{\xi}$Se$_2$   solar cell with: (i) a  CIGS layer that is nonhomogeneous
along the thickness direction and (ii) a  metallic   backreflector corrugated periodically along a fixed
direction. The bandgap in the CIGS layer was either sinusoidally or linearly graded.

A $27.7$\% efficiency,  $33.16$~mA~cm$^{-2}$ short-circuit current density, $1070$~mV open-circuit voltage, 
and  $78$\% fill factor can be achieved with a  2200-nm-thick CIGS layer that is sinusoidally graded
and is accompanied by a periodically corrugated Mo backreflector. There is no change in the foregoing data if the corrugations of the backreflector are suppressed.
In comparison, the efficiency is  $18.93$\%, the 
short-circuit current density is  $31.11$~mA~cm$^{-2}$,  the open-circuit voltage  is $742$~mV, and the fill factor
is $82$\%, when the bandgap is homogeneous and the backreflector is flat.

 In addition, a  $22.89$\% efficiency,  $29.18$~mA~cm$^{-2}$ short-circuit current density, $1045$~mV open-circuit voltage, and  $75$\% fill factor can be achieved with a 600-nm-thick CIGS layer that is sinusoidally nonhomogeneous and is accompanied by a periodically corrugated Mo backreflector. The efficiency reduces to $22.75$\% is the backreflector is flat, indicating the modest role of SPP waves  \cite{Anderson1983,Heine1995} and waveguide modes  \cite{Khaleque2013} for efficiency enhancement when the CIGS layer is ultrathin.
In comparison, the efficiency is  $13.79$\%, the short-circuit current density is  $24.66$~mA~cm$^{-2}$,  the open-circuit voltage  is $704$~mV, and the fill factor is $79$\% when the CIGS layer is homogeneous and the backreflector is flat. The periodic corrugation of the backreflector explains some of the enhancement of efficiency, but the majority of the enhancement is seen to be driven by the bandgap nonhomogeneity.
Efficiency enhancement can also be achieved by linearly grading the bandgap, but the gain is significantly smaller.
	
Optoelectronic optimization thus indicates that $22.89$\% efficiency is achievable  with ultrathin 
  solar cells with a 600-nm-thick CIGS layer. This efficiency compares favorably with the 22\% efficiency demonstrated with CIGS layers that are  more than three times thicker.  Thus, bandgap nonhomogeneity in conjunction with periodic corrugation of the backreflector can be effective in realizing ultrathin  CIGS solar cells  that can  help us in tackling the scarcity of indium  and provide a way towards  multi-terawatt  solar-power production. We hope that our model-predicted results will provide an impetus to devise efficacious techniques for bandgap grading of ultrathin CIGS layers.

\vspace{3mm}	
	
\noindent\textbf{Acknowledgments.} The authors thank two anonymous reviewers for invaluable suggestions to improve the contents of this paper. A. Lakhtakia thanks the Charles Godfrey Binder Endowment at the 
Pennsylvania State University for ongoing support of his research. The research of  F. Ahmad 
and A. Lakhtakia was partially supported by  US National Science Foundation (NSF) under grant 
number DMS-1619901. The research of  T.~H. Anderson and P.~B.  Monk was partially supported by  the US
NSF under grant number DMS-1619904.


\begin{thebibliography}{99}
		\bibitem{ZWS}
		P.~Jackson, R.~Wuerz, D.~Hariskos, E.~Lotter, W.~Witte, and M.~Powalla,
		``Effects of heavy alkali elements in Cu(In,Ga)Se$_2$ solar cells with efficiencies
		up to 22.6\%,"   {\textit {Phys. Status Solidi RRL}} {\bf 10}, 583--586 (2016).
		
		\bibitem{Green2018}
		M.~A.~Green, Y.~Hishikawa, E.~D.~Dunlop, D.~H.~Levi, J.~Hohl-Ebinger,
		and A.~W.~Y.~Ho-Baillie,
		``Solar cell efficiency tables (version 51)," \textit{Prog. Photovolt.: Res.
			Appl.} {\bf 26}, 3--12 (2018).
		
		
		\bibitem{Candelise_et_al-2012} 
		C.~Candelise, M.~Winskel, and R.~Gross,
		``Implications for CdTe and CIGS technologies
		production costs of indium and tellurium scarcity,"  
		{\textit {Prog. Photovolt: Res. Appl.}} {\bf 20}, 816--831 (2012).
		
		\bibitem{Gloeckler-Sites2005} 
		M.~Gloeckler and J.~R.~Sites, 
		``Potential of submicrometer thickness
		Cu(In,Ga)Se$_2$ solar cells,"  {\textit {J.  Appl. Phys.}} {\bf 98}, 103703 (2005).
		
		\bibitem{Schmid2017} 
		M.~Schmid, ``Review on light management by nanostructures in
		chalcopyrite solar cells,"  {\textit {Semicond. Sci. Technol.}} {\bf 32}, 043003 (2017).
		
		\bibitem{Schmid2015}  
		C.~van~Lare, G.~Yin, A. Polman, and M.~Schmid, 
		``Light coupling and trapping in ultrathin Cu(In,Ga)Se$_2$ 
		solar cells using dielectric scattering patterns,"
		{\textit{ACS Nano}} {\bf 9}(10), 9603--9613 (2015).
		
		\bibitem{P-Lalanne2017} 
		J.~Goffard, C.~Colin, F.~Mollica, A.~Cattoni, C.~Sauvan, P.~Lalanne,
		J.-F.~Guillemoles, N.~Naghavi, and S. Collin, 
		``Light trapping in ultrathin CIGS solar cells with
		nanostructured back mirrors,"  {\textit {IEEE J. Photovolt.}} {\bf 7}, 1433--1441 (2017).
		
		\bibitem{Vermang}
		B.~Vermang, J.~T.~W\"{a}tjen, V.~Fj\"allstr\"om, F.~Rostvall, M.~Edoff, 
		R.~Kotipalli, F.~Henry, and D.~Flandre, 
		``Employing Si solar cell technology to increase
		efficiency of ultra-thin Cu(In,Ga)Se$_2$ solar cells," 
		{\textit {Prog. Photovolt: Res. Appl.}} {\bf 22}, 1023--1029 (2014).

		\bibitem{Gloeckler-Sites2005JPCS} 
		M.~Gloeckler and J.~R.~Sites, ``Band-gap grading in Cu(In,Ga)Se$_2$ 
		solar cells,"  {\textit {J. Phys.   Chem.   Solids}} {\bf 66}, 1891-1894 (2005).	
		

		
		
		\bibitem{Songetal2004} 
		J. Song, S.~S. Li, C.~H. Huang, O.~D. Crisalle, and T.~J. Anderson, 
		``Device modeling and simulation of
		the performance of Cu(In$_{1-x}$,Ga$_x$)Se$_2$ solar cells,"
		{\textit{Solid-State Electron.}} {\bf 48}, 73--79 (2004).
		
		\bibitem{Song2010} 
		S.~H.~Song, K.~Nagaich, E.~S.~Aydil, R.~Feist, R.~Haley, and S.~A.~Campbell, 
		``Structure optimization for a high
		efficiency CIGS solar cell,"
		{\textit{Proc. 35th IEEE Photovolt. Special. Conf. (PVSC)}}, pp.~2488--2492,
		Honolulu, HI, USA, 20-25 June (2010).
		
		
		\bibitem{Anderson2017}
		T.~H.~Anderson, T.~G.~Mackay, and A.~Lakhtakia, 
		``Enhanced efficiency of 
		Schottky-barrier solar cell with periodically nonhomogeneous 
		indium gallium nitride layer,"
		{\textit{J. Photon. Energy}} {\bf 7}, 014502 (2017).
		
		\bibitem{Anderson2018}	
		T.~H.~Anderson, A.~Lakhtakia,	and P.~B.~Monk,
		``Optimization of nonhomogeneous
		indium-gallium-nitride Schottky-barrier
		thin-film solar cells," {\textit{J. Photon. Energy}} {\bf 8}, 034501 (2018).
		
		\bibitem{Faryad2013} 
		M.~Faryad and A.~Lakhtakia, 
		``Enhancement of light absorption efficiency of amorphous-silicon thin-film
		tandem solar cell due to multiple surface-plasmon-polariton 
		waves in the near-infrared spectral regime," 
		{\textit{Opt. Eng.}} {\bf 52}, 087106 (2013); errata: {\bf 53}, 129801 (2014).
		
		\bibitem{Liu2015} 
		L.~Liu, M.~Faryad, A.~S.~Hall, G.~D.~Barber, S.~Erten, T.~E.~Mallouk,
		A.~Lakhtakia, and T.~S.~Mayer, 
		``Experimental excitation of multiple 
		surface-plasmon-polariton waves and waveguide modes in a one-dimensional 
		photonic crystal atop a two-dimensional metal grating,"  
		{\textit{J. Nanophotonics}} \textbf{9}, 093593 (2015).
		
		\bibitem{Haug2011} 
		F.-J.~Haug, K.~S\"oderstr\"om, A.~Naqavi, and C.~Ballif,
		``Excitation of guided-mode resonances in thin film silicon solar cells,"
		{\textit{MRS Symp. Proc.}} {\bf 1321}, 123--128 (2011).
		
		\bibitem{Khaleque2013} 
		T.~Khaleque and R.~Magnusson, ``Light management through 
		guided-mode resonances in thin-film silicon solar cells," 
		\textit{J. Nanophotonics} \textbf{8}, 083995 (2014).
		
		\bibitem{Fonash} 
		S.~J.~Fonash, \emph{Solar Cell Device Physics} 
		(Academic Press,  2010).
		
		\bibitem{GG}
		E.~N.~Glytsis and T.~K.~Gaylord, ``Rigorous three-dimensional 
		coupled-wave diffraction analysis
		of single and cascaded anisotropic gratings," 
		{\textit{J. Opt. Soc. Am. A}} {\bf 4}, 2061--2080 (1987).
		
		\bibitem{ESW2013}  
		J.~A. Polo Jr., T.~G.~Mackay, and A.~Lakhtakia, 
		\emph{Electromagnetic Surface Waves:
			A Modern Perspective} (Elsevier,  2013).
			
		\bibitem{Ahmad2018}
		F.~Ahmad, T.~H.~Anderson, B.~J.~Civiletti, P.~B.~Monk, and A.~Lakhtakia, 
		``On optical-absorption peaks in a nonhomogeneous thin-film
		solar cell with a two-dimensional periodically corrugated metallic backreflector,"
		{\textit{J. Nanophotonics}} {\bf 12}, 016017 (2018).
		
		\bibitem{SSAM15G}
		National Renewable Energy Laboratory,
		\href{http://rredc.nrel.gov/solar/spectra/am1.5/}{Reference Solar Spectral Irradiance:
			Air Mass 1.5}
		(5 June 2018).
		
		\bibitem{Jenny_Book}
		J.~Nelson, {\em The Physics of Solar Cells} (Imperial College Press,  2003).
		
		
		
		\bibitem{Lehrenfeld}
		C.~Lehrenfeld, \textit{Hybrid Discontinuous Galerkin Methods for Solving
			Incompressible Flow Problems}, Diplomingenieur Thesis,
		Rheinisch-Westfa\"{a}lischen
		Technischen Hochschule, Aachen, Germany (2010).
		
		\bibitem{CockburnHDG}
		B.~Cockburn, J.~Gopalakrishnan, and R.~Lazarov, ``Unified hybridization of
		discontinuous Galerkin, mixed, and continuous Galerkin methods for second
		order elliptic problems,'' \textit{SIAM J. Numer. Anal.}~{\bf 47},
		1319--1365 (2009).
		
		\bibitem{Frisk14}
		C.~Frisk, C.~Platzer-Bj\"{o}rkman, J.~Olsson, P.~Szaniawski, J.~T. W\"{a}tjen,
		V.~Fj\"{a}llstr\"{o}m, P.~Salom\'e, and M.~Edoff, 
		``Optimizing Ga-profiles
		for highly efficient {Cu(In, Ga)Se$_2$} thin film solar cells in simple and
		complex defect models,'' 
		\textit{J. Phys. D: Appl. Phys.} {\bf 47}, 485104 (2014).

		 \bibitem{Kuciauskas2013}
		D. Kuciauskas, J. V. Li, M. A. Contreras, J. Pankow, P. Dippo, M. Young, L.
		M. Mansfield, R. Noufi, and D. Levi, ``Charge carrier dynamics and recombination
		in graded band gap CuIn$_{1-x}$Ga$_x$Se$_2$
		polycrystalline thin-film photovoltaic solar
		cell absorbers," 
		\textit{J. Appl. Phys.} {\bf 144}, 154505 (2013).
		
	

		\bibitem{DEA}
		R. Storn and K. Price, ``Differential evolution---a simple and efficient
		heuristic for global optimization over continuous spaces," {\textit {J. Global
				Optim.}} {\bf 11}, 341--359 (1997).
 		
		\bibitem{Ben2018} 
		B.~J.~Civiletti, T.~H.~Anderson, F.~Ahmad, P.~B.~Monk, and  
		A.~Lakhtakia, 
		``Optimization approach for optical absorption in
		three-dimensional structures including solar cells," 
		{\textit{Opt. Eng.}} {\bf 57}, 057101 (2018).
		
 		\bibitem{Ahmad-SPIE2018}  
		F.~Ahmad, T.~H.~Anderson, P.~B.~Monk, and A.~Lakhtakia,
		``Optimization of light trapping in ultrathin nonhomogeneous CuIn$_{1-\xi}$Ga$_{\xi}$Se$_2$ 
		solar cell backed by 1D periodically corrugated backreflector," {\textit{Proc. SPIE}}
		{\bf 10731}, 107310L  (2018).
 		
		\bibitem{mgf2}  
		M.~J. Dodge, ``Refractive properties of magnesium 
		fluoride,"  {\textit {Appl. Opt.}} {\bf 23}, 1980--1985 (1984).
		
		\bibitem{Rajan} 
		G.~Rajan, K.~Aryal, T.~Ashrafee, S.~Karki, A.-R.~Ibdah, V.~Ranjan, 
		R.~W.~Collins, and S. Marsillac, ``Optimization of anti-reflective coatings 
		for CIGS solar cells via real time spectroscopic ellipsometry," 
		{\textit{Proc. 42nd IEEE Photovolt. Special. Conf. (PVSC)}}, 
		New Orleans, LA, USA, 14--19 June (2015).
		
		\bibitem{AZO} 
		N.~Ehrmann and R.~Reineke-Koch, 
		``Ellipsometric studies on ZnO:Al thin
		films: Refinement of dispersion theories," 
		{\textit{Thin Solid Films}} {\bf 519}, 1475--1485 (2010).
		
		\bibitem{iZnO}	C. Stelling, C. R. Singh, M. Karg, T. A. F. K\"onig, 
		M. Thelakkat, M. Retsch, ``Plasmonic nanomeshes: their ambivalent 
		role as transparent electrodes in organic solar cells," {\textit{Sci. Rep.}} 
		{\bf{7}}, 42530 (2017). 			
		
			\bibitem{Jahagirdar2003} A. H. Jahagirdar, A. A. Kadam, and N. G. Dhere, ``Role of i-ZnO in optimizing open circuit voltage of CIGS$2$ and CIGS thin film solar cells," {\textit{Proc. 4th IEEE World Conf. on Photovolt. Energ.}}, Waikoloa, HI, USA, 7--12 May (2006).
			
		\bibitem{treharne}  
		R.~E.~Treharne, A.~Seymour-Pierce, K.~Durose, K.~Hutchings, S.~Roncallo, and
		D.~Lane, ``Optical design and fabrication of fully sputtered CdTe/CdS solar cells," 
		{\textit {J. Phys.: Conf. Ser.}} {\bf 286}, 012038	(2011).
	
		\bibitem{Al2O3}  
		R. Boidin, T. Halenkovi\u{c}, V. Nazabal, L. Bene\u{s}, and P. N\u{e}mec, ``Pulsed laser deposited alumina thin films," {\textit{Ceramics Int.}} {\bf42}, 1177--1182 (2016).
		
		\bibitem{Mo} 
		 M. R. Querry, ``Optical constants of minerals and other materials from the millimeter to the ultraviolet,"  \href{https://apps.dtic.mil/dtic/tr/fulltext/u2/a192210.pdf}{{\textit {Contractor Report}} CRDEC-CR-88009} (1987).
		
		\bibitem{Iskander}
		M.~F. Iskander, {\em Electromagnetic Fields and Waves}
		(Waveland Press,  
		2012).
		
		\bibitem{RJMP}
		R.~J.~Mart\'in-Palma and A.~Lakhtakia, \textit{Nanotechnology: A Crash Course}
		(SPIE, 2010).
	
		 \bibitem{Lindahl2013}	
		J. Lindahl, U. Zimmermann, P. Szaniawski, T. T\"{o}rndahl, A. Hultqvist, P. Salom\'{e}, C. Platzer-Bj\"{o}rkman, 
		and M. Edoff, ``Inline Cu(In,Ga)Se$_2$ co-evaporation for high-efficiency solar cells 
		and modules," \textit{IEEE J. Photovolt.} {\bf 3}, 1100--1105 (2013).

		\bibitem{Minoura} 
		S.~Minoura, T.~Maekawa, K.~Kodera, A.~Nakane, S. Niki,
		and H.~Fujiwara, 
		``Optical constants of Cu(In,Ga)Se$_2$ for arbitrary Cu and Ga compositions,"
		{\textit{J. Appl. Phys.}} {\bf 117}, 195703 (2015).

	
		
 		\bibitem{Brezzi2002}
		F.~Brezzi, L.~D. Marini, S.~Micheletti, P.~Pietra, R.~Sacco, and S.~Wang,
		``Discretization of semiconductor device problems ({I}),'' In: 
		W.~H.~A. Schilders and E.~J.~W. ter Maten (eds), 
		\textit{Handbook of Numerical Analysis: Numerical Methods for 
		Electrodynamic Problems}, 
		pp.~317--342 (Elsevier, 2005).
		
		\bibitem{Anderson-JCP}
		 T. H. Anderson, B. J. Civiletti, P. B. Monk, and A. Lakhtakia,
		``Coupled optoelectronic simulation and optimization of thin-film photovoltaic
		solar cells,"   
		\href{https://arxiv.org/abs/1906.03962}{arXiv: 1906.03962} (2019).

		
		\bibitem{FuQiuHDG}
		G.~Fu, W.~Qiu, and W.~Zhang, ``An analysis of {HDG} methods for
		convection-dominated diffusion problems," 
		{\em ESAIM: Math. Model. Numer. Anal.}~{\bf 49}, 225--256 (2015).
		
		
		\bibitem{Chen2016}
		Y.~Chen, P.~Kivisaari, M.-E.~Pistol, and N. Anttu,
		``Optimization of the short-circuit current in an InP nanowire array solar 
		cell through opto-electronic modeling," {\textit {Nanotechnology}} {\bf 27}(43),  
		435404 (2016).
		
		
		\bibitem{Brinkman}
		D.~Brinkman, K.~Fellner, P.~Markowich, and M.-T. Wolfram, ``A
		drift-diffusion-reaction model for excitonic photovoltaic bilayers:
		Asymptotic analysis and a 2--{D} {HDG} finite-element scheme," 
		{\em Math. Models Methods Appl. Sci.}~{\bf 23}, 839--872 (2013).
		
		\bibitem{abushama2005} J. AbuShama, R. Noufi, S. Johnston, S. Ward, and X. Wu, ``Improved performance in CuInSe$_2$ and surface-modified
		CuGaSe$_2$ solar cells," {\em Proc. 31st IEEE Photovolt. Special. Conf. (PVSC)}, pp. 299--302,
		Lake Buena Vista, FL, USA, 3--7 June (2005).
		
		
 		\bibitem{Anderson1983}
		L. M. Anderson, ``Harnessing surface plasmons for solar energy conversion," 
		\textit{Proc. SPIE} \textbf{408}, 172--178 (1983).
		
		\bibitem{Heine1995}
		C. Heine and R. F. Morf, ``Submicrometer gratings 
		for solar energy applications," \textit{Appl. Opt.}
		\textbf{34}, 2476--2482 (1995).
		
				
		\bibitem{Shoji_paper}
		A. S. Hall, M. Faryad, G. D. Barber, L. Liu, S. Erten, T. S. Mayer, 
		A. Lakhtakia, and T.  E. Mallouk,
		``Broadband light absorption with multiple surface plasmon 
		polariton waves excited at the interface of a metallic grating and photonic crystal,"
		\textit{ACS Nano} {\bf 7}, 4995--5007 (2013).
		


		\bibitem{Shuba2}
		M.~V. Shuba and A. Lakhtakia, ``Splitting of absorptance peaks in absorbing
		multilayer backed by a periodically
		corrugated metallic reflector," 
		{\textit{J. Opt. Soc. Am. A}} {\bf 33}, 779--784 (2016).
		
		
\end{thebibliography}
 \end{document}